\documentclass[10pt,journal,compsoc]{IEEEtran}
\ifCLASSOPTIONcompsoc
  \usepackage[nocompress]{cite}
\else
  \usepackage{cite}
\fi
\ifCLASSINFOpdf
   \usepackage[pdftex]{graphicx}
   \graphicspath{{../pdf/}{../jpeg/}}
   \DeclareGraphicsExtensions{.pdf,.jpeg,.png}
\else
   \usepackage[dvips]{graphicx}
   \graphicspath{{../eps/}}
   \DeclareGraphicsExtensions{.eps}
\fi

\usepackage{amsfonts}
\usepackage{booktabs}
\usepackage{multirow}
\usepackage{subfigure}
\usepackage{tabularx}
\usepackage{algorithm}
\usepackage{algorithmic}
\usepackage{amsmath}
\usepackage{amssymb}
\usepackage{color}
\usepackage{array}
\usepackage{hyphenat}
\begin{document}
%
\title{Hierarchical Multi-Agent Optimization for Resource Allocation in Cloud Computing}
%
%
%
%

\author{Xiangqiang~Gao,
        Rongke~Liu,~\IEEEmembership{Senior~Member,~IEEE}
        and~Aryan~Kaushik
\IEEEcompsocitemizethanks{\IEEEcompsocthanksitem X.~Gao and R.~Liu are with the School of Electronic and Information Engineering, Beihang University, Beijing, 100191.\protect\\
E-mail: \{xggao, rongke\_liu\}@buaa.edu.cn.
\IEEEcompsocthanksitem A.~Kaushik is with the School of Engineering, the University of Edinburgh, Edinburgh EH9 3JL, United Kingdom.\protect\\
E-mail: A.Kaushik@ed.ac.uk.}
}

\IEEEtitleabstractindextext{%
\begin{abstract}
In cloud computing, an important concern is to allocate the available resources of service nodes to the requested tasks on demand and to make the objective function optimum, i.e., maximizing resource utilization, payoffs and available bandwidth. This paper proposes a hierarchical multi-agent optimization (HMAO) algorithm in order to maximize the resource utilization and make the bandwidth cost minimum for cloud computing. The proposed HMAO algorithm is a combination of the genetic algorithm (GA) and the multi-agent optimization (MAO) algorithm. With maximizing the resource utilization, an improved GA is implemented to find a set of service nodes that are used to deploy the requested tasks. A decentralized-based MAO algorithm is presented to minimize the bandwidth cost. We study the effect of key parameters of the HMAO algorithm by the Taguchi method and evaluate the performance results. When compared with genetic algorithm (GA) and fast elitist non-dominated sorting genetic (NSGA-II) algorithm, the simulation results demonstrate that the HMAO algorithm is more effective than the existing solutions to solve the problem of resource allocation with a large number of the requested tasks. Furthermore, we provide the performance comparison of the HMAO algorithm with the first-fit greedy approach in on-line resource allocation.
\end{abstract}

}

\maketitle

\IEEEdisplaynontitleabstractindextext

%
\IEEEpeerreviewmaketitle

\IEEEraisesectionheading{\section{Introduction}\label{Introduction}}

%
%
%
%
\IEEEPARstart{F}{or} some electronic devices, which are composed of dedicated hardware equipments, i.e., field programmable gate array (FPGA), digital signal processor (DSP) and integrated circuit (IC), the compatibilities for different requested tasks are difficult to guarantee and the systems will be more complicated with the increase in the number of the requested tasks. The software defined network (SDN) and virtualization technology are the foundations of the cloud computing, and provide a promising and flexible approach to facilitate resource allocation \cite{kim2013improving,sezer2013we,erdogmus2009cloud}. Cloud service providers can allocate the available resources related to service nodes to the requested tasks depending on demand and supply. When a task consists of multiple sub-tasks, these sub-tasks could be deployed on several service nodes and form a service chain, which is a data flow through the service nodes in sequence and can be presented as a directed acyclic graph (DAG) \cite{moens2014vnf}. Each sub-task needs the physical resources for central processing unit (CPU), memory, or graphic processing unit (GPU). Besides, there are bandwidth costs to transfer data on different service nodes. For example, in case of data transmission, it includes five sub-tasks and the service chain about these sub-tasks can be represented as: network receiving $\rightarrow$ capture $\rightarrow$ tracking $\rightarrow$ synchronization $\rightarrow$ decoding, where each functional module is achieved by software programming and can run on a commonly used computer system. The complexity and development cost of a system can be effectively reduced by cloud computing, and the flexibility and scalability can also be improved. However, a new challenge in cloud computing is how to effectively allocate the available resources related to service nodes to the requested tasks, which leads to a combinatorial optimization problem \cite{kazmi2017hierarchical,zheng2017hybrid}.\par

\subsection{Literature Review}\label{Literature Review}

The optimization problems for resource allocation in cloud computing have been widely studied \cite{rankothge2017optimizing,qu2016delay,tseng2017dynamic,tan2017nsga,khebbache2018multi}, which are proved to be NP-hard and the complexities are analyzed in \cite{sun2016forecast,amaldi2016computational,addis2018complexity}. Meta-heuristic algorithms are effective optimization approaches for solving these resource allocation problems. Several variants of genetic algorithm (GA) are developed to improve the performance of the resource allocation solution in \cite{rankothge2017optimizing,qu2016delay,tseng2017dynamic} and the fast elitist non-dominated sorting genetic algorithm (NSGA-II), which is described in detail in \cite{deb2002fast}, is also used to tackle this problem \cite{tan2017nsga,khebbache2018multi}. Two modified particle swarm optimizations (MPSO) are proposed to reallocate the migrated virtual machines and achieve the resource management based on a flexible cost in \cite{dashti2016dynamic} and \cite{mani2017flexible}, respectively. Moreover, an ant colony optimization (ACO) for dealing with the nonlinear resource allocation problem is presented in \cite{yin2006ant}. To enhance the efficiency in terms of seeking the optimal solution, the authors in \cite{muthulakshmi2017hybrid} introduce a hybrid optimization algorithm of simulated annealing and artificial bee colony (ABC-SA).\par

However, as the scale of the optimization problem grows, a large feasible solution space needs to be searched and the computational complexity order of seeking the optimal solution increases. Hence, the performance of solving the problem could be reduced by using meta-heuristic algorithms \cite{omidvar2015designing,liu2001scaling}. To further solve this issue, some optimization algorithms based on decomposition and cooperative co-evolutionary method are introduced, where a large-scale problem is divided into several small-scale problems and global optimal solution can be obtained by addressing these sub-problems with cooperative co-evolutionary method \cite{wu2019new,ren2018boosting}. Reference \cite{yang2017Efficient} proposes a new cooperative co-evolution framework (CCFR), which can efficiently allocate computational resources based on the contributions of different sub-populations, to address a large-scale optimization problem.\par

Compared with centralized optimization methods, distributed optimization algorithms based on multi-agent systems (MAS) \cite{stone2000multiagent} have an explicit potential advantage for solving the task deployment and resource allocation in cloud computing \cite{kaihara2003multi,de2012multiagent,wang2016multiagent}. In \cite{kaihara2003multi}, the product allocation problem for supply chain market is considered as a discrete resource allocation and solved by a multi-agent based distributed optimization algorithm. In addition, an efficient greedy algorithm with multi-agent is proposed to address the task allocation problem in social networks \cite{de2012multiagent}, where the agents just require their local information about tasks and resources, and provide the resources for the tasks by an auction mechanism. Another auction-based virtual machine resource allocation approach with the multi-agent system is presented to save energy cost, and the virtual machines assigned on different agents can be exchanged by a local negotiation-based approach in \cite{wang2016multiagent}.\par

\subsection{Contributions}\label{Contributions}
In this paper, we assume that the information about the requested tasks and service nodes is given in advance, e.g., task types, the number of tasks, resource requirements for each sub-task and resource capacity values of service nodes. Furthermore, we formulate the problem of resource allocation to maximize the resource utilization and minimize the bandwidth cost under resource constraints. The resource utilization is defined as the ratio of the number of resources used by all the requested tasks to the total number of resources of the service nodes that are used to deploy the requested tasks. In order to address the optimization problem, we propose a hierarchical multi-agent optimization (HMAO) algorithm which is a combination of improved GA and multi-agent optimization (MAO) algorithm.\par

Firstly, we decompose the main objective of maximizing the resource utilization and minimizing the bandwidth cost into two sub-objectives: maximizing the resource utilization and minimizing the bandwidth cost \cite{yang2017Efficient}. The two sub-objective optimization problems are in conflict with each other and we assume that the former is considered with a higher priority. The improved GA is used to seek the optimal solution in order to maximize the resource utilization and the optimal solution can be expressed as a set of service nodes that are used to deploy the requested tasks. For the MAO algorithm, there are two types of agents: service agent and shared agent. Service agents are assigned to each service node to assist in resource management. A shared agent holds the information about resource allocation for all the service nodes and supports the service agents in \textit{access} and \textit{update} processes. The agents have environment-aware, autonomy, social behavior and load-balancing properties\cite{zheng2015multi}. The service agents visit the shared agent to obtain the information about resource allocation for all the service agents, and they can migrate and swap their sub-tasks with each other by selection and exchange operators that are designed based on a probabilistic approach. In addition, a priority-based source sub-task selection mechanism for the selection operator is implemented by considering load-balancing on the service nodes and the relationships between different sub-tasks. As a result, a global optimal solution will be obtained by seeking the optimal solutions of the service agents with cooperative co-evolutionary method \cite{ren2018boosting}. The proposed HMAO algorithm provides the following contributions.\par

\begin{enumerate}
  \item Considering to solve the joint problem of optimizing the resource utilization and bandwidth cost as an entire problem, it will increase the computational complexity of the optimization problem and weaken the performance of seeking the optimal solution, especially, for the high-dimensional problems. In this paper, we decompose the total optimization problem into two sub-problems. Correspondingly, a hierarchical multi-agent optimization algorithm, which combines an improved GA with MAO algorithm, is presented for solving the two optimization sub-problems. So that we can effectively reduce the computational complexity of the overall optimization problem.

  \item The set of available service nodes, which represents the optimal solution of maximizing the resource utilization, can be obtained by the improved GA. Then the MAO algorithm is proposed to solve the sub-problem of minimizing the bandwidth cost. We consider four main characteristics for the agents: environment-aware, autonomy, social behavior and load-balancing. Furthermore, we design the action sets and behavior criteria for service agents and a shared agent, the relationships between different agents are illustrated based on an organized architecture. To migrate and swap those sub-tasks on the service nodes, we implement two operators of selection and exchange, where the selection operator consists of a source sub-task selection and a target sub-task selection. Considering load-balancing on the service nodes and the relationships between different sub-tasks, the source sub-task is provided by a priority-based selection mechanism. For the MAO algorithm, a feasible solution is partitioned into several small-scale solutions and each service agent indicates one part. We can find the global optimal solution by optimizing the objectives of the service agents with cooperative co-evolutionary method.

  \item To keep diversity in feasible solutions and avoid premature convergence, the selection and exchange operators for the MAO algorithm are implemented based on a probabilistic method. For the former, we can randomly choose a sub-task from the service nodes as the target sub-task through the selection probability. Similarly, if the objective result for a service agent does not improve after the exchange, it can also continue to be executed with the use of exchange probability. By introducing a probabilistic method to the selection and exchange operators, we can further improve the performance of the proposed HMAO algorithm.
\end{enumerate}
\begin{figure}[tbp]
  \centering
  \includegraphics[width=0.36\textwidth]{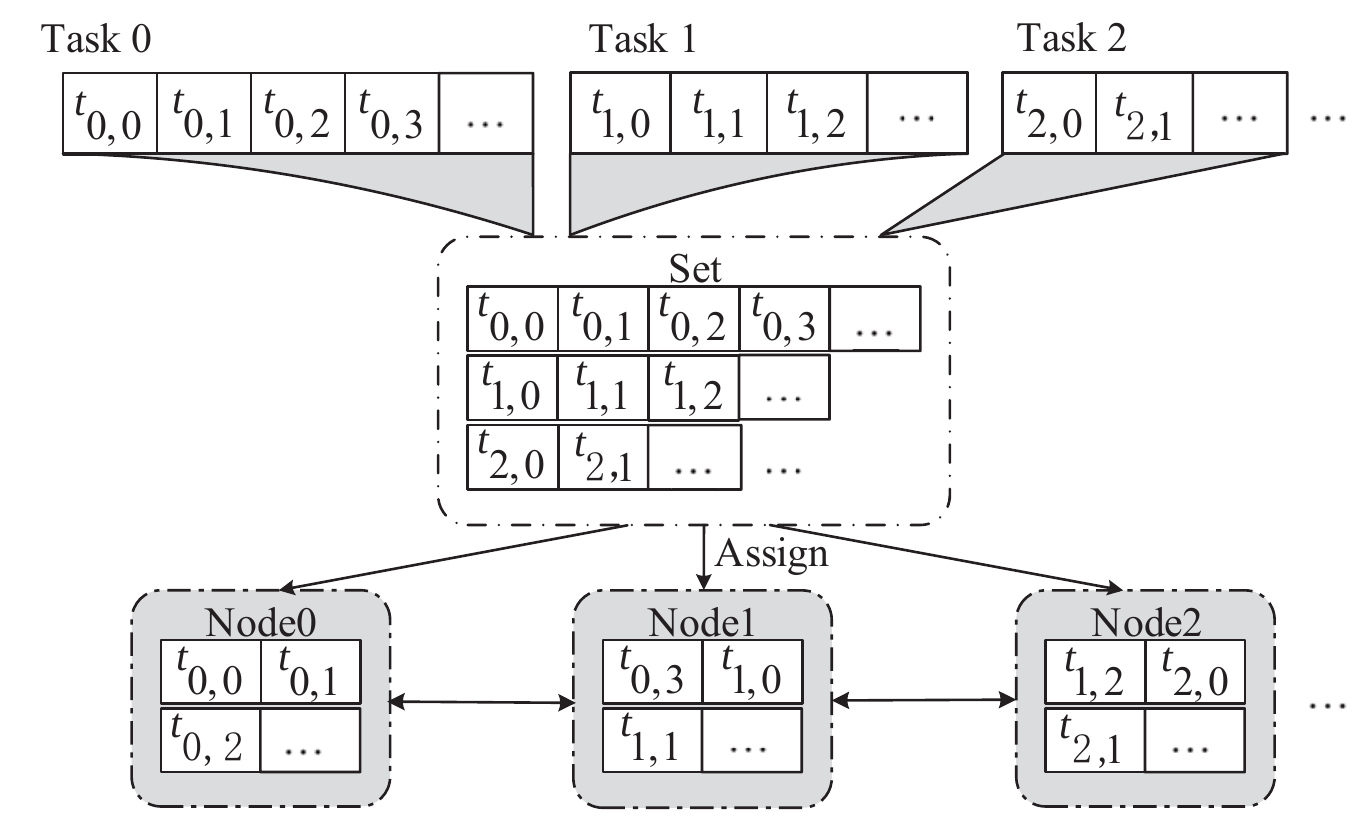}
  \caption{Procedure for resource allocation.}
  \label{Procedure for resource allocation}
\end{figure}
Finally, several experiments for different tasks are carried out to verify the performance of the proposed HMAO algorithm. The results show that the proposed HMAO algorithm is an effective approach to solve the optimization problem of resource allocation. When compared with GA and NSGA-II, the proposed HMAO algorithm performs better for high-dimensional problems in terms of solution quality, convergence time and stability.\par

The remainder of this paper is organized as follows. Section \ref{System Model} introduces the system model of resource allocation. In Section \ref{Problem Formulation}, we provide the problem formulation for the system model. A hierarchical multi-agent optimization algorithm is proposed in Section \ref{Hierarchical Multi-agent Optimization}. Section \ref{Performance Evaluation} investigates the effect of key parameters for the proposed HMAO algorithm and evaluates the performance comparison with existing optimization algorithms. The conclusion of this paper is discussed in Section \ref{Conclusion}.\par

\section{System Model}\label{System Model}

A cloud computing system can be modeled as a graph $G = <V,E>$, where $V=\left \{v_0,v_1,\dots,v_{K-1} \right \}$ represents a set of service nodes with $K$ service nodes, $v_i$ is the $i$-th service node. $E$ describes a set of links between these service nodes and $<v_i,v_j>\in E$ indicates the link between service nodes $v_i$ and $v_j$. Service node $v_i$ contains four resources: CPU, memory, GPU and bandwidth, whose capacities are denoted as $C^i,M^i,G^i$ and $B^i$, respectively. We also assume that any two service nodes can communicate with each other through inter-connected networks, i.e., $\forall <v_i,v_j>\in E$. Let $T=\left \{t_0,t_1,\dots,t_{L-1} \right \}$ be a set of tasks with $L$ total number of tasks and $t_l$ denotes the $l$-th task. $t_l$ consists of $N_l$ sub-tasks and is expressed as $t_l=\left \{ t_{l,0},t_{l,1},\dots,t_{l,N_l-1} \right \}$, which is an ordered list and there is a precedence relationship between different sub-tasks. That is, $t_{l,n}$ can not be carried out until all its predecessors are finished. The resource requirements of CPU, memory, GPU and bandwidth for $t_{l,n}$ running on $v_i$ are described as $c_{l,n}^i,m_{l,n}^i,g_{l,n}^i,b_{l,n}^i$, respectively. In cloud computing, an effective resource allocation approach is used to allocate the available resources of service nodes to the requested tasks. Note that the total number of resources on any service node can not be more than its capacities, and the bandwidth cost will be considered when two adjacent sub-tasks are placed on different service nodes. Moreover, we can deploy as many sub-tasks as possible to a service node in order to improve the resource utilization \cite{rankothge2017optimizing,sun2016forecast,khebbache2018multi}.\par

\begin{figure}[tbp]
  \centering
  \subfigure[]{\includegraphics[width=0.23\textwidth]{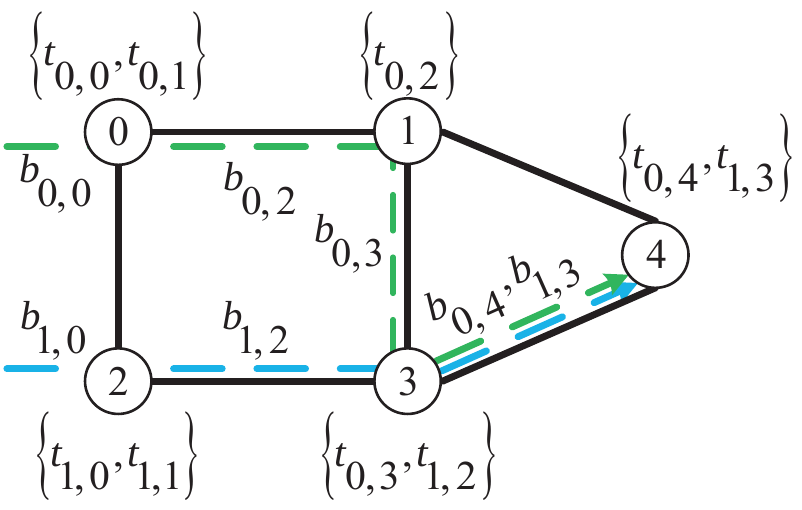}
  \label{Example of resource allocation with different schemes-a}}
  \subfigure[]{\includegraphics[width=0.23\textwidth]{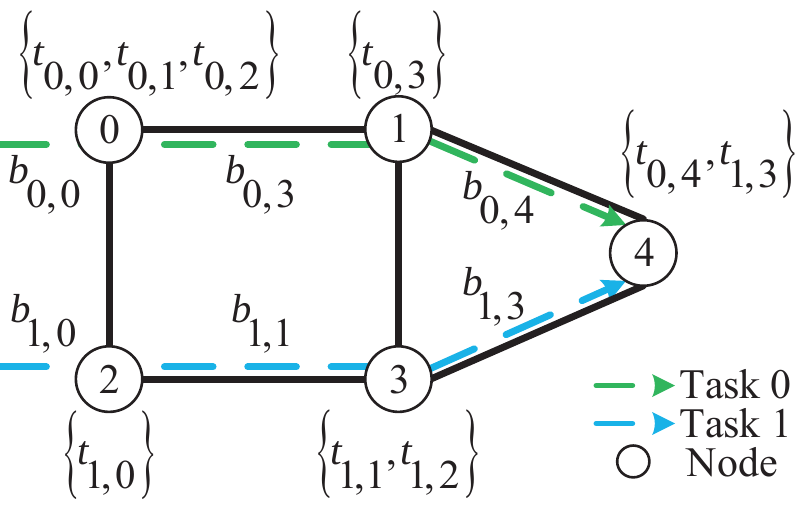}
  \label{Example of resource allocation with different schemes-b}}
  \caption{Example of resource allocation with different schemes.}
  \label{Example of resource allocation with different schemes}
\end{figure}

The procedure for resource allocation in our system model is shown in Fig.~\ref{Procedure for resource allocation}, where the information about the requested tasks and service nodes is obtained in advance, all sub-tasks from the requested tasks are deployed to multiple service nodes simultaneously, and the neighbouring sub-tasks can be located on the same service node or different ones. Therefore, different resource allocation schemes have an impact on the system performance \cite{sun2016forecast}.\par

Fig.~\ref{Example of resource allocation with different schemes} presents an example of resource allocation with different schemes. There are five service nodes with communication links in the network, two tasks are given as $t_0=\left \{ t_{0,0},t_{0,1},t_{0,2},t_{0,3},t_{0,4}\right \}$ and $t_1=\left \{ t_{1,0},t_{1,1},t_{1,2},t_{1,3}\right \}$. In Fig.~\ref{Example of resource allocation with different schemes-a}, for task $t_{0}$, sub-tasks $t_{0,0},t_{0,1}$ are deployed on service node $v_0$, sub-tasks $t_{0,2},t_{0,3}$ and $t_{0,4}$ are on service nodes $v_1,v_3$ and $v_4$, respectively. A service chain for $t_0$ is built on the service nodes $v_0,v_1,v_3$ and $v_4$ in order, and there is no bandwidth cost between sub-tasks $t_{0,0}$ and $t_{0,1}$ as they are on the same service node. Hence, the bandwidth cost for $t_0$ is $b_{0,0}^0+b_{0,2}^1+b_{0,3}^3+b_{0,4}^4$. For task $t_1$, sub-tasks $t_{1,0}$ and $t_{1,1}$ are allocated on service node $v_2$, and sub-tasks $t_{1,2}$ and $t_{1,3}$ are on service nodes $v_3$ and $v_4$, respectively. The bandwidth cost for $t_1$ is indicated as $b_{1,0}^2+b_{1,2}^3+b_{1,3}^4$. Different resource allocation schemes for tasks $t_0$ and $t_1$ are described in Fig.~\ref{Example of resource allocation with different schemes-b}. sub-tasks $t_{0,0},t_{0,1}$ and $t_{0,2}$ are assigned on service node $v_0$, and sub-tasks $t_{0,3}$ and $t_{0,4}$ are on service nodes $v_1$ and $v_4$, respectively, and the bandwidth cost for $t_0$ is $b_{0,0}^0+b_{0,3}^1+b_{0,4}^4$. Similarly, the bandwidth cost for $t_1$ is $b_{1,0}^2+b_{1,1}^3+b_{1,3}^4$. As shown in Fig.~\ref{Example of resource allocation with different schemes}, the bandwidth costs vary for different resource allocation schemes.

\section{Problem Formulation}\label{Problem Formulation}

\begin{table}[htbp]
  \renewcommand{\arraystretch}{1.3}
  \caption{List of Symbols}
  \label{List of Symbols}
  \centering
  \resizebox{\columnwidth}{!}{
  \begin{tabular}{l|l}
  \hline
  \bfseries Symbol & \bfseries Definition\\
  \hline\hline
  $V$ & Set of service nodes in cloud computing.\\
  $E$ & Set of network links in cloud computing.\\
  $K$ & Maximum number of service nodes.\\
  $V_a$ & Set of service nodes deployed tasks.\\
  $K_a$ & Number of service nodes deployed tasks.\\
  $T$ & Set of tasks.\\
  $L$ & Number of tasks.\\
  $N_l$ & Number of sub-tasks for the $l$-th task.\\
  $t_l$ & Set of sub-tasks for the $l$-th task.\\
  $t_{l,n}$ & The $n$-th sub-task for the $l$-th task.\\
  $v_i$ & The $i$-th service node. \\
  $C^i$ & Resource capacity for CPU on $v_i$.\\
  $M^i$ & Resource capacity for Memory on $v_i$.\\
  $G^i$ & Resource capacity for GPU on $v_i$.\\
  $B^i$ & Resource capacity for Bandwidth on $v_i$.\\
  $c_{l,n}^i$ & Resource requirement of $t_{l,n}$ for CPU on $v_i$.\\
  $m_{l,n}^i$ & Resource requirement of $t_{l,n}$ for Memory on $v_i$.\\
  $g_{l,n}^i$ & Resource requirement of $t_{l,n}$ for GPU on $v_i$.\\
  $b_{l,n}^i$ & Resource requirement of $t_{l,n}$ for Bandwidth on $v_i$.\\
  $x_{l,n}^i$ & Indicate whether $t_{l,n}$ is assigned on $v_i$. \\
  $y_{l,n,\hat{n}}^{i,j}$ & Indicate whether a bandwidth cost is available for $t_{l,n},t_{l,\hat{n}}$. \\
  $\tilde{b}_{l,n}^{i}$ & Bandwidth cost for running $t_{l,n}$.\\
  $U_{l,n}^{s}$ & Set of successors of $t_{l,n}$.\\
  $U_{l,n}^{p}$ & Set of predecessors of $t_{l,n}$.\\
  $Z_{1}$ & Objective function of resource utilization.\\
  $Z_{2}$ & Objective function of bandwidth cost.\\
  $Z$ & Total objective function.\\
  $\alpha,\beta$ & Weight values.\\
  \hline
  \end{tabular}
  }
\end{table}
In this section, a mathematical description of resource allocation is presented for our system model. Our purpose is to maximize the resource utilization and minimize the bandwidth cost in cloud computing with several physical constraints. Moreover, we do not consider the network resource constraints in this paper, such as routers and switches. The main symbols used to formulate our problem are summarized in Table~\ref{List of Symbols}.\par

To further discuss the problem, we define a binary decision variable $x_{l,n}^i$ that indicates whether sub-task $t_{l,n}$ is deployed on service node $v_i$, i.e., $x_{l,n}^i = 1$ means $t_{l,n}$ is allocated on $v_i$, otherwise not. In addition, another binary decision variable $y_{l,n,\hat{n}}^{i,j}$ is used to describe whether there is a bandwidth cost between sub-tasks $t_{l,n}$ and $t_{l,\hat{n}}$. Let us denote a set of predecessors of $t_{l,n}$ as $U_{l,n}^{p}$ and a set of successors of $t_{l,\hat{n}}$ as $U_{l,\hat{n}}^{s}$. We assume that $t_{l,n}$ and $t_{l,\hat{n}}$ are placed on $v_i$ and $v_j$, respectively. When $U_{l,n}^{p} \neq \varnothing, t_{l,\hat{n}} \in U_{l,n}^{p}$, then $y_{l,n,\hat{n}}^{i,j} = 1$, or else $y_{l,n,\hat{n}}^{i,j} = 0$. Note that the bandwidth cost between a sub-task and the source node is not neglected. The bandwidth cost for $t_{l,n}$ can be written as:

\begin{equation} \label{equation1}
\tilde{b}_{l,n}^{i} =\begin{cases}
b_{l,n}^{i}, & \text{if } U_{l,n}^{p}=\varnothing \quad \text{or} \quad y_{l,n,\hat{n}}^{i,j} = 1,  \\
0, & \text{otherwise}.
\end{cases}
\end{equation}\par

With the physical resource constraints \cite{tseng2017dynamic}, the number of resources used by the requested tasks on a service node should be less than the resource capacities. As a result, the resource constraints about CPU, memory and GPU are given as follows:

\begin{equation} \label{equation2}
\begin{cases}
\sum\limits_{l=0}^{L-1}\sum\limits_{n=0}^{N_l-1}c_{l,n}^{i}x_{l,n}^{i} \leq C^{i},\\
\sum\limits_{l=0}^{L-1}\sum\limits_{n=0}^{N_l-1}m_{l,n}^{i}x_{l,n}^{i} \leq M^{i},\\
\sum\limits_{l=0}^{L-1}\sum\limits_{n=0}^{N_l-1}g_{l,n}^{i}x_{l,n}^{i} \leq G^{i}.\\
\end{cases}
\end{equation}\par

\noindent Similarly, as the amount of bandwidth used on a service node can not be more than the capacity, the bandwidth resource constraint is expressed as:

\begin{equation} \label{equation3}
\sum\limits_{l=0}^{L-1}\sum\limits_{n=0}^{N_l-1} \tilde{b}_{l,n}^{i}x_{l,n}^{i} \leq B^{i}.
\end{equation}\par

\noindent Furthermore, all the sub-tasks from the task set $T$ are deployed to the service nodes in $V$ and any sub-task can be allocated just only once. Thus, we can obtain the following:

\begin{equation} \label{equation4}
\sum\limits_{i=0}^{K-1}\sum\limits_{l=0}^{L-1}\sum\limits_{n=0}^{N_l-1}x_{l,n}^{i} = \sum\limits_{l=0}^{L-1}N_{l}.
\end{equation}\par

In this paper, one of our goals is to improve the resource utilization of cloud computing by reducing the number of service nodes used. For service node $v_i$, the resource utilization includes three parts: CPU utilization, memory utilization and GPU utilization, which are denoted by $\varphi_c^i,\varphi_m^i,\varphi_g^i$, respectively. They can be computed as follows:

\begin{equation} \label{equation5}
\begin{cases}
\varphi_c^i = \sum\limits_{l=0}^{L-1}\sum\limits_{n=0}^{N_l-1}c_{l,n}^{i}x_{l,n}^{i}/C^i,\\
\varphi_m^i = \sum\limits_{l=0}^{L-1}\sum\limits_{n=0}^{N_l-1}m_{l,n}^{i}x_{l,n}^{i}/M^i,\\
\varphi_g^i = \sum\limits_{l=0}^{L-1}\sum\limits_{n=0}^{N_l-1}g_{l,n}^{i}x_{l,n}^{i}/G^i.\\
\end{cases}
\end{equation}\par

According to the preferences of different resource types, we provide the weight values for CPU, memory and GPU as $\alpha _c,\alpha _m$ and $\alpha _g$, respectively. The resource utilization $Z_{1}$ for our system model is obtained by a linear weighted sum method as follows:

\begin{equation} \label{equation6}
Z_{1}(X) =\frac{1}{K_a}\sum\limits_{i=0}^{K_a-1}\left ( \alpha _c\varphi_c^i + \alpha _m\varphi_m^i + \alpha _g\varphi_g^i\right ),
\end{equation}\par

\noindent where $X=\left \{ x_{l,n}^{i},\forall v_i \in V,\forall t_{l,n} \in T \right \}$ is a feasible solution for allocating the available resources of service nodes to the requested tasks in cloud computing. The parameter $K_a$ $(K_a \leq K)$ denotes the number of service nodes that are used to deploy the requested tasks. In addition, $\alpha _c +\alpha _m + \alpha _g = 1$.\par

\noindent Another goal is to optimize the bandwidth cost by deploying the adjacent sub-tasks to the same service node. Let us denote the bandwidth cost as $Z_{2}$ which can be expressed as follows:

\begin{equation} \label{equation7}
Z_{2}(X) = \frac{1}{K_a}\sum\limits_{i=0}^{K_a-1}\sum\limits_{l=0}^{L-1}\sum\limits_{n=0}^{N_l-1} \tilde{b}_{l,n}^{i}x_{l,n}^{i}.
\end{equation}\par

Next, we need to simultaneously optimize these two objectives which are $Z_{1}(X)$ and $Z_{2}(X)$. Specifically, our purpose is to seek an optimal solution which maximizes the resource utilization $Z_{1}(X)$ and minimizes $Z_{2}(X)$. Thus the combined objective $Z(X)$ which involves maximizing $Z_{1}(X)$ and minimizing $Z_{2}(X)$ can be expressed as:

\begin{equation} \label{equation8}
 \textrm{maximize} \quad Z(X) = \beta_{1}Z_{1}(X) + \beta_{2}(1-Z_{2}(X)),
\end{equation}\par

\noindent where $\beta_{1}$ and $\beta_{2}$ are the weight values for $Z_{1}(X)$ and $1-Z_{2}(X)$, respectively. The preferences of different objectives can be adjusted by varying $\beta_{1}$ and $\beta_{2}$, and we consider $\beta_{1} + \beta_{2} = 1$. However, this optimization problem is regarded as NP-hard problem, which means a high computational complexity order for finding an optimal solution. Meta-heuristic optimization algorithms are effective approaches to solve the combinatorial optimization problem, but the performance of solution quality and convergence time will degrade with the increase in the scale of the problem. Distributed optimization algorithms, which are based on decomposition and multi-agent systems, can improve the optimization solution effectively.\par
In the next section, we propose the HMAO algorithm, which is an optimization approach using hierarchical multi-agent framework, to address the resource allocation problem in cloud computing.\par

\begin{figure}[tbp]
  \centering
  \includegraphics[width=0.36\textwidth]{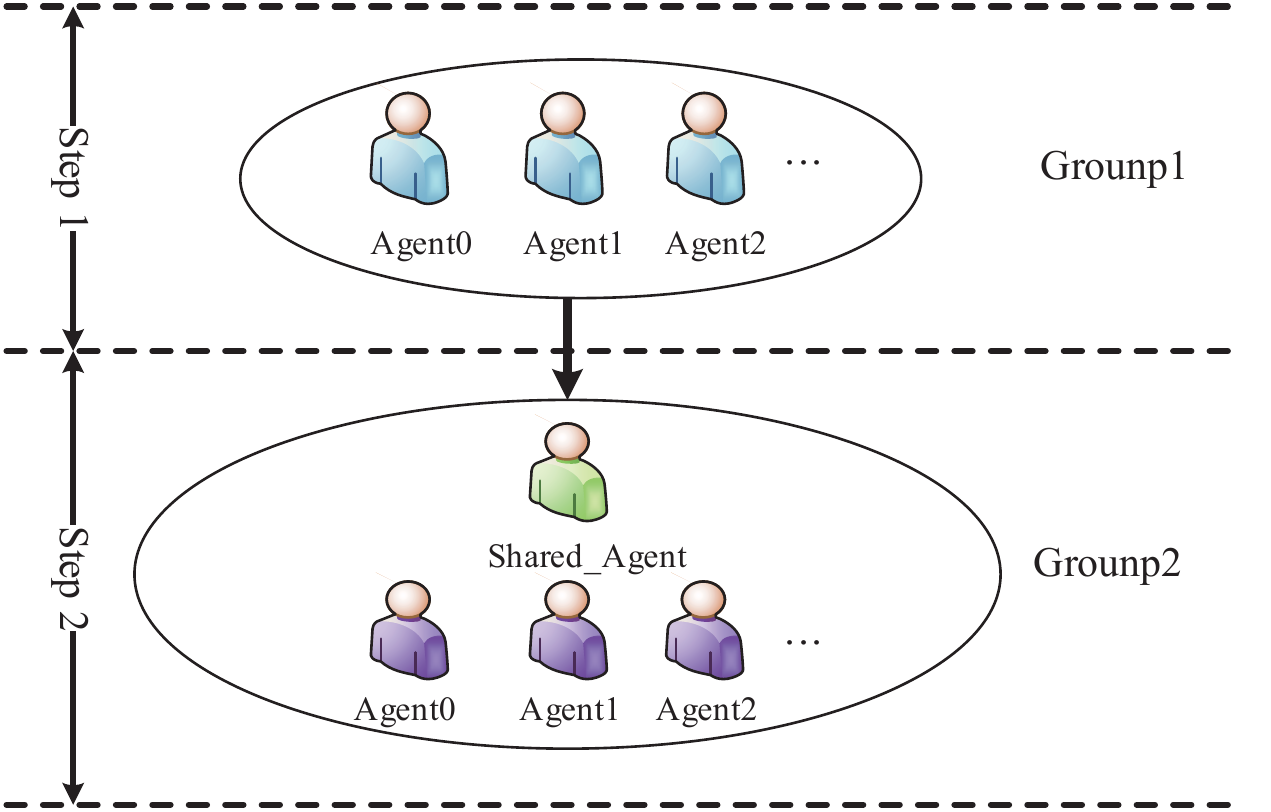}
  \caption{Proposed hierarchical multi-agent framework.}
  \label{Proposed hierarchical multi-agent framework}
\end{figure}

\section{Hierarchical Multi-agent Optimization}\label{Hierarchical Multi-agent Optimization}

In this paper, we address a joint optimization problem of maximizing the resource utilization in cloud computing systems and reducing the bandwidth cost. To reduce the computational complexity of the optimization problem, we decompose the total objective into two optimization sub-problems: maximizing $Z_{1}(X)$ and minimizing $Z_{2}(X)$, and these two optimization sub-problems will be solved, accordingly.\par

Firstly, an improved GA is introduced to find the optimal solution which maximizes $Z_{1}(X)$. All sub-tasks from $T$ make up an ordered list as an individual that represents a feasible solution. We use a roulette wheel selection approach \cite{razali2011genetic} to obtain those individuals with higher fitness values. Besides, two-point crossover and signal-point mutation are also used \cite{hartmann1998competitive}. A set $V_a$ of service nodes that are used to deploy the requested tasks can be provided as the optimal solution by the improved GA.\par

Then, a multi-agent optimization algorithm is proposed to solve the sub-problem of minimizing $Z_{2}(X)$. We use a shared agent to hold the information of resource allocation for the service nodes, and assign service agents to each service node for assisting in resource management. Different service agents can cooperate, coordinate and compete with each other to optimize their objectives with respect to the behavior criteria \cite{wang2016multiagent}. To keep the diversity of feasible solutions and avoid the occurrence of premature convergence, both selection and exchange operators are achieved by a probabilistic method. In addition, a feasible solution consists of all the sub-tasks from the available service agents and those sub-tasks on a service agent are considered as part of a feasible solution. We can optimize the objectives of the service agents with cooperative co-evolutionary method to obtain a global optimal solution \cite{ren2018boosting}.\par

\begin{figure}[tbp]
  \centering
  \includegraphics[width=0.36\textwidth]{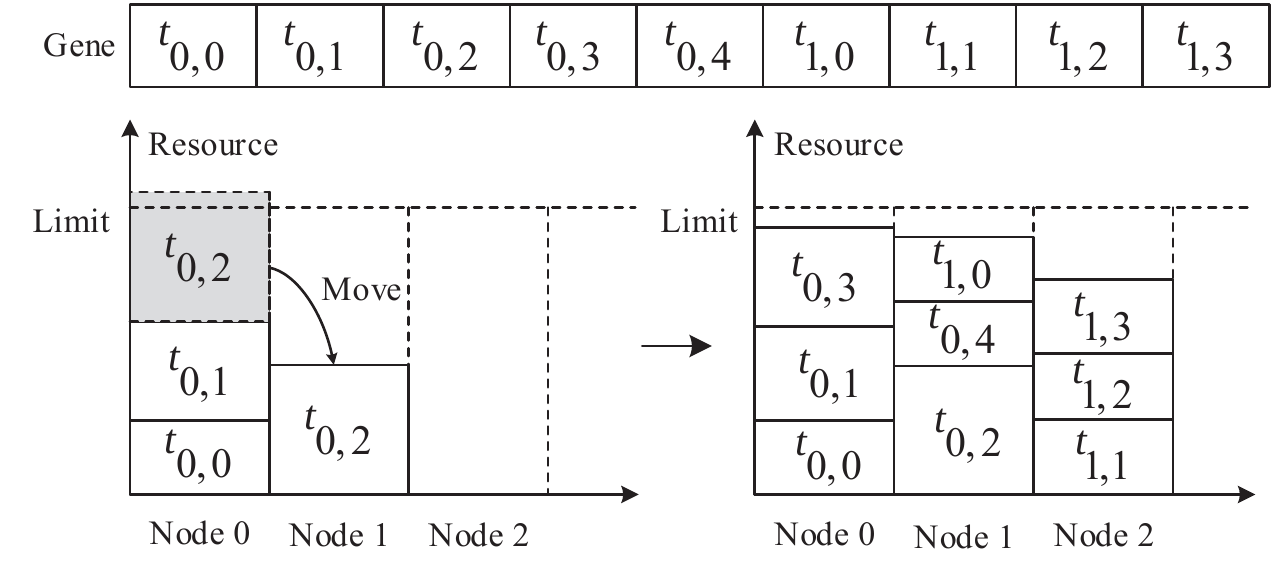}
  \caption{Procedure for decoding an individual.}
  \label{Procedure for decoding an individual}
\end{figure}

Fig.~\ref{Proposed hierarchical multi-agent framework} shows the proposed hierarchical multi-agent framework. The procedure for addressing the optimization problem is divided into two steps. In the first step, the improved GA is used to find the optimal solution which maximizes $Z_{1}(X)$, and each agent represents an individual. In the second step, the MAO algorithm is applied to seek the optimal solution which minimizes $Z_{2}(X)$.\par

\subsection{Improved GA}

For the sub-problem of optimizing the resource utilization, the objective is to maximize $Z_{1}(X)$ and the constraints need to be satisfied. The improved GA is introduced to solve this optimization problem.\par

In the improved GA, a population includes $P$ individuals, an individual $p \in P$ is encoded by a permutation representation \cite{hartmann1998competitive,kaur2012efficient} and consists of $\sum\limits_{l=0}^{L-1}N_l$ genes in order, where a gene represents a sub-task. During the process of decoding, all service nodes in $V$ are sorted in ascending order by index, and the service nodes with low index have high priority to be deployed with the requested tasks. Thus, a sub-task is deployed depending upon the priority of a service node until the resource requirements are satisfied. By the decoding method, the sub-tasks from an individual can be allocated to these service nodes in $V$ in sequence, and the result of resource allocation indicates a feasible solution.\par

Fig. \ref{Procedure for decoding an individual} illustrates the procedure for decoding an individual. There are two tasks: $t_0$ with 5 sub-tasks and $t_1$ with 4 sub-tasks, an individual is expressed as $p= \left \{ t_{0,0},t_{0,1},t_{0,2},t_{0,3},t_{0,4},t_{1,0},t_{1,1},t_{1,2},t_{1,3}\right \}$. Three service nodes are denoted as $v_0,v_1$ and $v_2$, respectively, and their resource capacities are limited. Firstly, it is seen that sub-tasks $t_{0,0}$ and $t_{0,1}$ are allocated to service node $v_0$, but the required resources for $v_0$ are more than the capacities after deploying $t_{0,2}$ to $v_0$. As a result, sub-task $t_{0,2}$ is moved to service node $v_1$. Due to the high priority of service nodes with low index, sub-task $t_{0,3}$ will be placed to service node $v_0$. Similarly, we deploy sub-tasks $t_{0,4}$ and $t_{1,0}$ to service node $v_1$, sub-tasks $t_{1,1},t_{1,2}$ and $t_{1,3}$ to service node $v_2$.\par

Three operators for selection, crossover and mutation are used as follows:
\begin{itemize}
  \item \emph{Selection operator:} A roulette wheel selection \cite{razali2011genetic} is applied to obtain the individuals with high fitness values. All individuals are sorted in ascending order by their fitness values and the cumulative probability distribution function (CDF) is computed accordingly. Then we choose the candidates through the concept of the survival of the fittest,which means an individual with high fitness is more likely to be chosen.\par

  \item \emph{Crossover operator:} Two-point crossover \cite{hartmann1998competitive} is used as the crossover operator and executed with crossover probability $p_c$. We randomly select two gene points for two individuals (one each from the mother and father) in the field $\left \{1, \sum\limits_{l=0}^{L-1}N_l-2 \right \}$ to mate with each other, each offspring inherits some of the genes from their parents, respectively. An example of two-point crossover is described in Fig. \ref{Example of two-point crossover operator}. Two individuals $p_0= \left \{ t_{0,3},t_{0,1},t_{0,4},t_{1,0},t_{0,0},t_{1,1},t_{0,2},t_{1,2},t_{1,3} \right \},p_1= \left \{ t_{0,0},t_{0,4},t_{1,3},t_{0,1},t_{1,2},t_{1,0},t_{0,3},t_{1,1},t_{0,2} \right \}$ are given as the parents, and two gene points are randomly generated, such as $3$ and $6$. Firstly, an offspring inherits $t_{0,3},t_{0,1}$ and $t_{0,4}$ from $parent0$ as the genes of itself. Then $3$ genes ($point 1 -point 0 =3$) from $parent1$ need to be searched from low to high, and they can not be the same as that of the offspring. Therefore, genes $t_{0,0},t_{1,3}$ and $t_{1,2}$ from $parent1$ are inherited as the genes of the offspring. In the same way, the offspring can inherit genes $t_{1,0},t_{1,1}$ and $t_{0,2}$ from $parent0$. Thus the offspring is indicated as $\left \{ t_{0,3},t_{0,1},t_{0,4},t_{0,0},t_{1,3},t_{1,2},t_{1,0},t_{1,1},t_{0,2}\right \}$. Similarly, $\left \{ t_{0,0},t_{0,4},t_{1,3},t_{0,3},t_{0,1},t_{1,0},t_{1,2},t_{1,1},t_{0,2} \right \}$ can represent another offspring.\par

        \begin{figure}[tbp]
          \centering
          \includegraphics[width=0.36\textwidth]{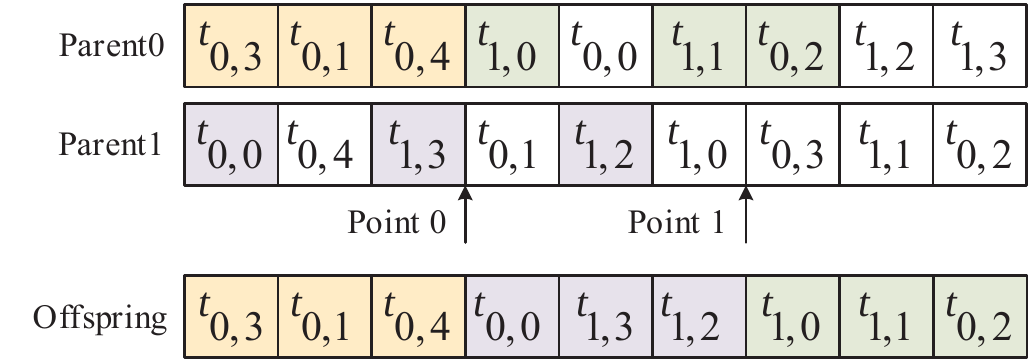}
          \caption{Example of two-point crossover operator.}
          \label{Example of two-point crossover operator}
        \end{figure}

  \item \emph{Mutation operator:} Mutation operator \cite{hartmann1998competitive} is carried out with mutation probability $p_m$, where we randomly choose two gene points from an individual and exchange their genes to gemerate a new individual.\par
\end{itemize}

\begin{algorithm}
  \caption{Improved GA.}
  \label{Improved GA}
  \begin{algorithmic}[1]
  \STATE \textbf{Initialize:} Population $P$, individual $p \in P$, crossover probability $p_c$, mutation probability $p_m$, maximum number of iterations $Iter_{max}$;
  \FOR {$Iter = 0$ to $Iter_{max}$}
  \STATE Compute the fitness for all individuals $P$;
  \STATE Obtain the CDF based on the fitness values;
  \STATE Run selection operator to get $P$ candidates;
  \FOR{$i = 0$ to $\frac{length(P)}{2}$}
  \STATE Randomly choose two individuals;
  \STATE Generate a random number $p_r$;
  \IF{$p_r \leq p_c$}
  \STATE Run two-point crossover operator;
  \ENDIF
  \ENDFOR
  \STATE Produce offspring population;
  \FOR{$i = 0$ to $length(P)$}
  \STATE Randomly choose an individual;
  \STATE Generate a random number $p_r$;
  \IF{$p_r \leq p_m$}
  \STATE Run mutation operator;
  \ENDIF
  \ENDFOR
  \STATE Update population $P$;
  \ENDFOR
  \end{algorithmic}
\end{algorithm}

The improved GA is described in Algorithm \ref{Improved GA}. Let us denote the maximum number of iterations as $Iter_{max}$. At the beginning, the initial population $P$ is randomly produced, we compute the fitness of the objective function for those individuals from population $P$ and obtain the CDF. Then the population of offsprings can be obtained by running selection, crossover and mutation operators, respectively. The optimal solution of maximizing $Z_{1}(X)$ will be given in iterative evolution. The termination criterion is met when the number of iterations is greater than $Iter_{max}$.\par

According to the improved GA, we obtain an optimal solution which maximizes the resource utilization, where the optimal solution indicates a set $V_a$ of service nodes that are used to deploy the requested tasks and contains $K_a$ service nodes. However, the bandwidth cost and load-balancing are not considered for this sub-problem. The following section \ref{Multi-agent Optimization} will discuss the sub-problem of minimizing the bandwidth cost and load-balancing based on a set $V_a$ of service nodes.\par

\subsection{Multi-agent Optimization}\label{Multi-agent Optimization}

For set $V_a$ of service nodes, we propose a multi-agent optimization approach to address the optimization sub-problem of minimizing $Z_{2}(X)$ with the constraints. For the MAO, the agents have four characteristics: environment-aware, autonomy, social behavior and load-balancing \cite{zheng2015multi}, which are described in detail as follows:\par

\begin{enumerate}
  \item \emph{Environment-aware:} The environment in the MAO algorithm consists of all the agents and their relationships, where the agents are designed as a shared agent and $K_a$ service agents in advance, and we use an organized architecture to illustrate their relationships. A service agent can obtain the information about resource allocation by accessing the shared agent, and interact with other service agents to better adapt to the current environment condition, that is, to achieve a higher fitness value. A shared agent provides access services for the service agents in order to help with resource management.\par

  \item \emph{Autonomy:} When the environment conditions are changed, an agent can autonomously make a decision for the next actions to adjust its fitness value by a set of actions and the behavior criteria. For a service agent, its action set includes four parts: access, selection, exchange and update. Firstly, it obtains the information about resource allocation by accessing the shared agent. Then selection and exchange operators, which are described later in this section, can be executed, respectively. The results of interaction on service agents will be updated on the shared agent. In addition, the action set of a shared agent can provide three services which are storing, accessing and updating the information.\par

  \item \emph{Social behavior:} According to social behavior, the agents share the information about resource allocation, and the sub-tasks deployed on different service agents can be exchanged with each other. There are two kinds of social behavior which are between shared agent and service agents, and service agents with each other. For the former, the agents can share the information about resource allocation with all the service agents. For the latter, the aim is to exchange those sub-tasks that are deployed on different service agents to improve their objectives.\par

  \item \emph{Load-balancing:} Load-balancing is an important issue to ensure that the service nodes are being used sufficiently. Therefore, considering load-balancing, a service agent runs the exchange operator by cooperating, coordinating and competing with other service agents.\par
\end{enumerate}

\begin{figure}[tbp]
  \centering
  \includegraphics[width=0.36\textwidth]{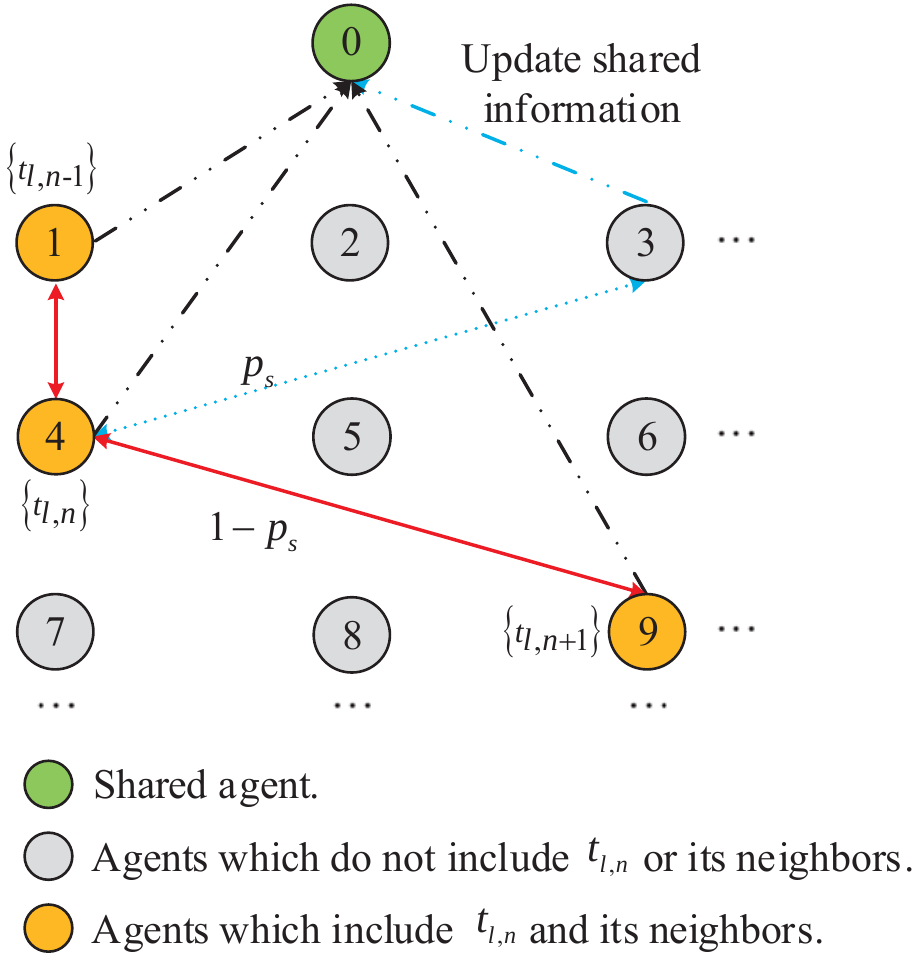}
  \caption{Architecture for multi-agent optimization.}
  \label{Architecture for multi-agent optimization}
\end{figure}

In order to better describe the MAO algorithm, some important specifications are given as follows:
\begin{itemize}
  \item \emph{Shared agent:} Shared agent, which is denoted by $a_s$, holds the information about task deployment and resource allocation for all service agents, it can support the service agents to access and update the information in real-time.

  \item \emph{Service agent:} Service agents are assigned to the service nodes in $V_a$ to assist in resource management, where they can access and update the information about resource allocation on the shared agent. Moreover, the sub-tasks deployed on different service agents can be migrated and swapped with each other by selection and exchange operators. The service agent assigned to $v_i$ is indicated as $a_i$ and all the service agents make up a set $A_a$ of the service agents.

  \item \emph{Adjacent agent:} We assume that $\forall t_{l,n} \in a_i$, $\exists t_{l,\hat{n}} \in a_j, i\neq j$, where $t_{l,\hat{n}} \in U_{l,n}^{p} \cup U_{l,n}^{s}$, then $a_i,a_j$ are seen as adjacent agents.

  \item \emph{Active sub-task:} For a sub-task $\forall t_{l,n} \in a_i$, its adjacent sub-tasks are migrated and swapped to $a_i$ to improve the bandwidth cost by selection and exchange operators. Sub-task $t_{l,n}$ is defined as an active sub-task.

  \item \emph{Host service agent:} If an active sub-task $t_{l,n} \in a_i$, the service agent $a_i$ is considered as the host service agent.

  \item \emph{Feasible solution:} All sub-tasks placed on service agent $a_i$ are indicated as $\Lambda_i$, where $\Lambda_i$ is considered as part of a feasible solution. Therefore, a feasible solution can be expressed as $X = \bigcup\limits_{i=0}^{K_a-1} \Lambda_i$.

  \item \emph{Objective function:} A feasible solution is decomposed into $K_a$ parts, thus the objective can be re-written as: $Z_{2}(X) = \frac{1}{K_a}\sum\limits_{i=0}^{K_a-1}Z_{2}(\Lambda_i)$.

\end{itemize}

In the MAO algorithm, there is a shared agent $a_s$ and $K_a$ service agents, the agents can communicate with each other by network links, such as $a_s$ and $a_i$, and $a_i$ and $a_j, i \neq j$. The service agents can share the information about resource allocation by accessing the shared agent $a_s$, migrate and swap their sub-tasks with each other to reduce their bandwidth costs. To keep the diversity of feasible solutions, selection and exchange operators are implemented based on a probabilistic method which is discussed later in this section.\par
\begin{figure}[tbp]
  \centering
  \includegraphics[width=0.36\textwidth]{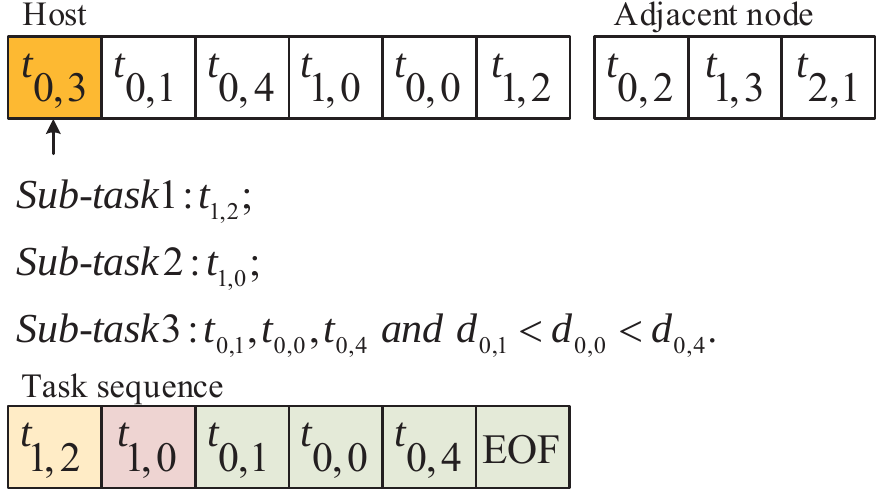}
  \caption{Source sub-task candidate list.}
  \label{Source sub-task candidate list}
\end{figure}
Generally, the procedure of the MAO includes four parts: obtain the adjacent agents, choose source and target sub-tasks, execute the exchange operator, and update the shared information. Firstly, let us select sub-task $\forall t_{l,n} \in a_i$ as the active sub-task, all its adjacent sub-tasks $U_{l,n}=U_{l,n}^{p} \cup U_{l,n}^{s}$ and their service agents can be obtained by accessing the shared agent. Then selection and exchange operators can be carried out for each sub-task from corresponding $U_{l,n}$. For the selection operator, we need to choose the source and target sub-tasks, where the target sub-task is selected by a probabilistic method, and the source sub-task from $a_i$ is obtained by a priority-based selection mechanism. For exchange operator, we can migrate and swap the source and target sub-tasks among different service agents with multiple constraints. Furthermore, the exchange operation can be carried out with exchange probability $p_e$ if the objective fitness is not improved at this instance. When the exchange is finished, the shared information will be updated.\par

An example of the MAO algorithm is illustrated in Fig.~\ref{Architecture for multi-agent optimization}. We assume that a host service agent is $a_4$ and an active sub-task is $t_{l,n}$. The service agent $a_4$ can obtain the information about the adjacent sub-tasks for $t_{l,n}$ by visiting the shared agent, and there are two adjacent sub-tasks $t_{l,n-1} \in a_1,t_{l,n+1} \in a_9$. For sub-task $t_{l,n+1}$, it is chosen as the target sub-task with selection probability $1-p_s$ or else a random sub-task $t_{\hat{l},\hat{n}} \in a_3$ is chosen to be the target sub-task. A source sub-task from $a_4$ is selected by a priority-based selection mechanism. Then we can run the exchange operation and update the shared information. Similarly, we can proceed with the processing for $t_{l,n-1}$.\par

Next, we discuss three main parts for the MAO algorithm.\par

\subsubsection{Target Selection}

In the MAO algorithm, to optimize the bandwidth cost, our purpose is to assign the adjacent sub-tasks from a task  to the same service agent as far as possible. In addition, a probabilistic method is used to select the target sub-task to avoid premature convergence.\par

For active sub-task $t_{l,n} \in a_i$, the set of its adjacent sub-tasks is $U_{l,n}$, and we assume that a candidate adjacent sub-task is $t_{l,\hat{n}} \in U_{l,n}, t_{l,\hat{n}} \in a_j, i \neq j$ and a random sub-task is denoted as $t_{\hat{l},\hat{n}} \in a_k, i \neq k$. Therefore, the target sub-task is $t_{\hat{l},\hat{n}}$ with probability $p_s$, otherwise $t_{l,\hat{n}}$. The algorithm for target selection is described in Algorithm \ref{Target selection}.\par
\begin{algorithm}[tbp]
  \caption{Target selection.}
  \label{Target selection}
  \begin{algorithmic}[1]
  \STATE \textbf{Initialize:} Probability $p_s$;
  \STATE For $\forall t_{l,n} \in a_i,\exists t_{l,\hat{n}} \in U_{l,n},t_{l,\hat{n}} \in a_j,a_i,a_j \in A_a,i \neq j$, $t_{\hat{l},\hat{n}} \in a_k, i \neq k$;
  \STATE Generate a random number $p_r$;
  \IF{$p_r \geq p_s$}
    \STATE Let $t_{l,\hat{n}}$ be the target sub-task;
  \ELSE
    \STATE Let $t_{\hat{l},\hat{n}}$ be the target sub-task;
  \ENDIF
  \end{algorithmic}
\end{algorithm}
\subsubsection{Source Selection}

Considering load-balancing and objective optimization, a priority-based source sub-task selection mechanism is proposed. In the host service agent, all available sub-tasks for active sub-task $t_{l,n} \in a_i$ are sorted in descend order by the priority, which includes two categories: the dependence relationships of sub-tasks for the host and adjacent service agents, and the used resources. There are three sub-task types according to the dependence relationships as follows:

\begin{itemize}
  \item \emph{Sub-task1:} Let us denote active sub-task as $t_{l,n} \in a_i$, an adjacent service agent as $a_j, i\neq j$ and a set of available sub-tasks as $W_{l,n}$. If $t_{{l}',{n}'} \in W_{l,n}, {t}'_{{l}',{n}'} \in U_{{l}',{n}'}$, make $\exists {t}'_{{l}',{n}'} \notin W_{l,n}$ and ${t}'_{{l}',{n}'} \in a_j$, let $t_{{l}',{n}'}$ be sub-task1.

  \item \emph{Sub-task2:} If $t_{{l}',{n}'} \in W_{l,n}, {t}'_{{l}',{n}'} \in U_{{l}',{n}'}$, make $\forall {t}'_{{l}',{n}'} \notin W_{l,n}$ and $ {t}'_{{l}',{n}'} \notin a_j$, let $t_{{l}',{n}'}$ be sub-task2.

  \item \emph{Sub-task3:} If $t_{{l}',{n}'} \in W_{l,n}, {t}'_{{l}',{n}'} \in U_{{l}',{n}'}$, make $\exists {t}'_{{l}',{n}'} \in W_{l,n}$, let $t_{{l}',{n}'}$ be sub-task3.

\end{itemize}\par

\noindent The precedence relations for three sub-task types are ranked as: sub-task1 $>$ sub-task2 $>$ sub-task3. Furthermore, for the same sub-task type, we calculate the resource utilization difference between $a_i$ and the average value of the system for each candidate sub-task. All the candidate sub-tasks are sorted in ascending order by the differences.\par

Fig. \ref{Source sub-task candidate list} describes the procedure of a source sub-task candidate list. Host service agent is indicated as $\left \{ t_{0,3},t_{0,1},t_{0,4},t_{1,0},t_{0,0},t_{1,2}\right \}$, active sub-task is $t_{0,3}$ and an adjacent service agent consists of $t_{0,2},t_{1,3}$ and $t_{2,1}$. We can find that $t_{1,2}$ is sub-task1, $t_{1,0}$ is sub-task2, $t_{0,1},t_{0,0}$ and $t_{0,4}$ are sub-task3. For sub-task3, the differences of the sub-tasks are ranked as $d_{0,1} < d_{0,0} < d_{0,4}$. As a result, the task list can be indicated as $\left \{ t_{1,2},t_{1,0},t_{0,1},t_{0,0},t_{0,4}\right \}$.\par

For the sub-task candidate list, the source sub-task can be chosen by the priority, and different source sub-tasks have an influence on the performance. The algorithm for source selection is shown in Algorithm \ref{Source selection}.\par

\begin{algorithm}[tbp]
  \caption{Source selection.}
  \label{Source selection}
  \begin{algorithmic}[1]
  \STATE For active sub-task $t_{l,n}\in a_i$, an adjacent service agent $a_j,i\neq j$;
  \STATE Make a set of available sub-tasks $W_{l,n}$;
  \FOR{$\forall t_{{l}',{n}'} \in W_{l,n}$}
    \STATE Obtain the set of its adjacent sub-tasks $U_{{l}',{n}'}$;
    \STATE For ${t}'_{{l}',{n}'} \in U_{{l}',{n}'}$;
    \IF{$\exists {t}'_{{l}',{n}'} \notin W_{l,n}$ and ${t}'_{{l}',{n}'} \in a_j$}
        \STATE Let $t_{{l}',{n}'}$ be sub-task1;
    \ELSIF{$\forall {t}'_{{l}',{n}'} \notin W_{l,n}$ and ${t}'_{{l}',{n}'} \notin a_j$}
        \STATE Let $t_{{l}',{n}'}$ be sub-task2;
    \ELSE
        \STATE Let $t_{{l}',{n}'}$ be sub-task3;
    \ENDIF
    \STATE Assume $t_{{l}',{n}'}$ is a source sub-task and calculate the difference between the resource utilization of $a_i$ and the average value of the system;
  \ENDFOR
  \STATE Sort the sub-task candidate list by precedence.
  \end{algorithmic}
\end{algorithm}
\subsubsection{Exchange Procedure}

For active sub-task $t_{l,n}$, the target sub-task and source sub-task candidate list are given by target and source selection operations, then the service agents can cooperate, coordinate and compete with each other to migrate and swap their sub-tasks to improve the objectives through the exchange operator. In this context, there are four situations to be considered as follows:

\begin{itemize}
  \item \emph{Objective optimization:} For the exchange procedure, our aim is to migrate and swap these sub-tasks on different service agents to reduce their bandwidth costs. That is, the objective fitness should be improved for the host service agent after running the exchange operator.

  \item \emph{Resource constraints:} The number of resources used for each service agent can not exceed its resource capacities.

  \item \emph{Load-balancing:} Load-balancing is implemented by a priority-based source sub-task selection mechanism and the procedure of migrating and swapping the sub-tasks.

  \item \emph{Diversity of feasible solutions:} Exchange operator is achieved by a probabilistic method, when the objective optimization is not satisfied, the exchange procedure can continue to be carried out with a lower probability.
\end{itemize}\par

\noindent Next, we describe the exchange procedure in detail.\par

Let us denote the source sub-task list by $Q_{l,n}$. For $\forall t_{l',n'} \in Q_{l,n}$, we firstly ensure that the objective result is improved by the exchange operation. If it is not, the exchange procedure can keep running with exchange probability $p_e$. Considering the load-balancing for all the service agents, if the load-balancing constraint in equation \eqref{equation9} is satisfied, we will just migrate the target sub-task from $a_j$ to $a_i$ and there will be nothing to do for the source sub-task. Besides, the migration will be also considered when there is no available source sub-task in the candidate list. The load-balancing constraint can be expressed as:\par

\begin{equation} \label{equation9}
\eta_{i} < \eta_{j} + 2*\bar{\eta},
\end{equation}\par

\noindent where $\eta_{i}$ and $\eta_{j}$ show the resource utilization of $a_i$ and $a_j$ after the migration, respectively. The parameter $\bar{\eta}$ indicates the average resource utilization for all sub-task types. In most cases, we need to swap the target and source sub-tasks with each other between $a_i$ and $a_j$. Note that the migration and exchange are satisfied with the resource constraints. The exchange procedure is shown in Algorithm \ref{Exchange procedure}.\par

\begin{algorithm}[tbp]
  \caption{Exchange procedure.}
  \label{Exchange procedure}
  \begin{algorithmic}[1]
  \STATE For target sub-task $t_{\hat{l},\hat{n}}$, source sub-task candidate list $Q_{l,n}$, exchange probability $p_e$;
  \FOR{$\forall t_{l',n'} \in Q_{l,n}$}
    \IF {The objective optimization is invalid}
        \STATE Generate a random number $p_r$;
        \IF {$p_r > p_e$}
            \STATE Continue;
        \ENDIF
    \ENDIF
    \IF{($t_{l',n'} = EOF$ or ($t_{l',n'} \neq EOF$ and hold equation \eqref{equation9})) and meet with resource constraints}
        \STATE Run the migration and break;
    \ELSIF{$t_{l',n'} \neq EOF$ and hold resource constraints}
        \STATE Run the exchange and break;
    \ENDIF
  \ENDFOR
  \STATE Update the shared information on the shared agent.
  \end{algorithmic}
\end{algorithm}

The MAO algorithm is an iterative optimization algorithm based on multi-agent systems and Algorithm \ref{MAO algorithm} describes the entire MAO algorithm. We assume that the maximum number of iterations is $M$. During an iterative evolution, all the sub-tasks from each service agent run the selection and exchange operations, and the objective value for this service agent can be improved with high probability. All the service agents can work together with cooperative co-evolutionary method. Thus, the global optimal solution can be found by increasing the number of iterations.\par
\begin{algorithm}[tbp]
  \caption{MAO algorithm.}
  \label{MAO algorithm}
  \begin{algorithmic}[1]
  \STATE \textbf{Input:} Maximum number of iterations $M$, selection probability $p_s$, exchange probability $p_e$;
  \FOR{$m = 0$ to $M$}
    \FOR{$\forall a_i \in A_a$}
        \FOR{$\forall t_{l,n} \in a_i$}
            \STATE Obtain $U_{l,n}$ by visiting $a_s$;
            \FOR{$\forall t_{l,\hat{n}} \in U_{l,n}, t_{l,\hat{n}} \in a_j, i\neq j$}
                \STATE Run the target selection operation;
                \STATE Run the source selection operation;
                \STATE Run the exchange procedure;
            \ENDFOR
        \ENDFOR
        \STATE Update $\Lambda_i$ for $a_i$;
    \ENDFOR
    \STATE Get the feasible solution $X=\bigcup\limits_{i=0}^{K_a-1}\Lambda_i$ and compute the objective fitness value;
  \ENDFOR
  \STATE Obtain an approximate optimal solution.
  \end{algorithmic}
\end{algorithm}
\subsection{HMAO Algorithm in On-line Resource Allocation}

\begin{algorithm}[tbp]
  \caption{HMAO Algorithm in on-line resource allocation.}
  \label{HMAO Algorithm in on-line resource allocation}
  \begin{algorithmic}[1]
  \STATE \textbf{Input at time $t$:} Set $T_{new,t} \in T$ of the tasks coming at $t$, set $T_{old,t-1} \in T$ of the tasks ending at $t-1$;
  \STATE Release the required resources of the old tasks in $T_{old,t-1}$ and update the resource information for the service nodes;
  \STATE Generate the initial population $P$ for $T_{new,t}$;
  \STATE Run the GA and obtain the set $V_{a,t}$ of service nodes that are used to deploy the new tasks;
  \STATE Carry out the MAO algorithm for $V_{a,t}$;
  \STATE Obtain an approximate optimal solution of allocating the available resources to $T_{n,t}$.
  \STATE Time: $t\leftarrow t+1$.
  \end{algorithmic}
\end{algorithm}

The proposed HMAO algorithm is easily to be applied to the on-line resource allocation by a few modifications. In this paper, we consider the scenario of allocating the available resources related to service nodes to the requested tasks in batch mode. That is, the requested tasks need to be performed are collected and will be handled at a fixed time slot. In order to improve the quality of the initial solutions for the GA, an individual is encoded by the inter-dependent relationships of those sub-tasks in sequence and the rest of the population are randomly encoded to keep diversity of feasible solutions. For the decoding procedure of the GA, we sort the service nodes that are used to deploy the requested tasks in ascending order by the resource utilization. All the sub-tasks for an individual are assigned to the available service nodes with the first-fit rule \cite{bhamare2017optimal,Espling2016modeling} and a new service node would be activated when any of the available service nodes can not meet the resource requirements of the sub-task to be assigned. In each time slot, some new requested tasks appear and several old requested tasks are over. Therefore, we can release the resources used by the old tasks and re-assign them to the new requested tasks. Algorithm \ref{HMAO Algorithm in on-line resource allocation} describes the proposed HMAO algorithm in on-line resource allocation. For time slot $t$, let us denote the set of the new requested tasks as $T_{new,t} \in T$ and the set of the old requested tasks that are ending as $T_{old,t-1} \in T$. Firstly, we end the old tasks from $T_{old,t-1}$ and release the required resources used. Then with existing resources used and physical resource constrains, we use the proposed HMAO algorithm to allocate the available resources of service nodes to the new requested tasks.\par

\section{Performance Evaluation}\label{Performance Evaluation}

In this section, we make several experiments for different number of communication tasks to verify the performance of the proposed HMAO algorithm with computer simulation results. Meanwhile, the performance of the proposed HMAO algorithm is analyzed by comparing with two existing baseline algorithms, which are GA and NSGA-II. The experimental platform is a high performance server, which is i7-4790k CPU, 16 GB memory and windows 10. Each communication task contains 5 parts: network receiving, capture, tracking, synchronization and decoding, and their resource requirements can be observed in Table \ref{Resource requirements for a task}. All service nodes are homogeneous and the resource capacities for CPU, memory, GPU and bandwidth are 2900 MHz, 96 GB, 8 and 1000 Mbps, respectively. Note that the limitation of network links is not considered in this paper, such as switches and routers. Equation \eqref{equation8} is considered for performance metrics comparison between the proposed HMAO algorithm and other existing baseline algorithms.\par
\begin{table}[tbp]
\centering
\caption{Resource requirements for a task.}
\label{Resource requirements for a task}
\resizebox{\columnwidth}{!}{
\begin{tabular}{|p{0.12\textwidth}<{\centering}|p{0.06\textwidth}<{\centering}|p{0.06\textwidth}<{\centering}|p{0.04\textwidth}<{\centering}|p{0.08\textwidth}<{\centering}|}
\hline
Name              & \begin{tabular}[c]{@{}c@{}}CPU\\(MHz)\end{tabular} & \begin{tabular}[c]{@{}c@{}}Memory\\(GB)\end{tabular} & GPU & \begin{tabular}[c]{@{}c@{}}Bandwidth\\(Mbps)\end{tabular} \\ \hline
Network receiving & 290                                                 & 9.6                                                   & 0                                                    & 100                                                        \\ \hline
Capture           & 319                                                 & 11.52                                                 & 1                                                    & 97                                                         \\ \hline
Tracking          & 435                                                 & 12.48                                                 & 1                                                    & 95                                                         \\ \hline
Synchronization   & 638                                                 & 12.48                                                 & 1                                                    & 92                                                         \\ \hline
Decoding          & 145                                                 & 4.8                                                   & 1                                                    & 90                                                         \\ \hline
\end{tabular}
}
\end{table}

\subsection{Simulation Parameters Setup}\label{Simulation Parameters Setup}
\begin{table}[tbp]
\centering
\caption{Parameters for Taguchi method.}
\label{Parameters for Taguchi method}
\resizebox{\columnwidth}{!}{%
\begin{tabular}{|p{0.07\textwidth}<{\centering}|p{0.07\textwidth}<{\centering}|p{0.07\textwidth}<{\centering}|p{0.07\textwidth}<{\centering}|p{0.07\textwidth}<{\centering}|}
\hline
\multirow{2}{*}{Factor} & \multicolumn{4}{c|}{Level} \\ \cline{2-5}
                        & 1     & 2    & 3    & 4    \\ \hline
$p_s$                      & 0.01  & 0.05 & 0.10 & 0.15 \\ \hline
$p_e$                      & 0.01  & 0.05 & 0.10 & 0.15 \\ \hline
$M$              & 250   & 500  & 750  & 1000 \\ \hline
\end{tabular}%
}
\end{table}
Firstly, we assume that the weights in equation \eqref{equation8} have the same values, and can be described as $\alpha _c = \alpha _m = \alpha _g = \frac{1}{3}$ and $\beta_{1} = \beta_{2} = \frac{1}{2}$. The parameters for the improved GA are $P = 16,Iter_{max}=5,p_c=1.0,p_m=0.1$. In the MAO algorithm, there are three main parameters: selection probability $p_s$, exchange probability $p_e$ and the number of iterations, which have impact on the performance results. In order to better estimate this impact on the performance for different combinations of parameters, the Taguchi method of design-of-experiment (DOE) is used to generate the test cases and analyze the results of our experiments \cite{zheng2015multi}. There are 3 factors and each factor contains 4 levels, and the combinations of different parameters for Taguchi method are shown in Table \ref{Parameters for Taguchi method}. Moreover, we develop the orthogonal table $L_{16}(4^3)$, where there are 16 cases and each case is carried out 10 times for 8, 16, 24, 32 tasks, respectively, to obtain the average results. Table \ref{Orthogonal table and the approximate solutions} provides the orthogonal table $L_{16}(4^3)$ and the values related to the approximate solutions for different tasks.

\begin{table}[tbp]
\centering
\caption{Orthogonal table $L_{16}(4^3)$ and the optimal solutions for $Z(X)$. }
\label{Orthogonal table and the approximate solutions}
\resizebox{\columnwidth}{!}{%
\begin{tabular}{|c|c|c|c|c|c|c|c|}
\hline
\multirow{2}{*}{No.} & \multicolumn{3}{c|}{Factor} & \multicolumn{4}{c|}{$Z(X)$} \\ \cline{2-8}
                     & $p_s$    & $P_e$    & $M$  & $L=8$  & $L=16$  & $L=24$  & $L=32$ \\ \hline
0                    & 0.01  & 0.01  & 250         & 0.7829  & 0.8008   & 0.8103   & 0.8109  \\ \hline
1                    & 0.01  & 0.05  & 500         & 0.7853  & 0.8032   & 0.8141   & 0.8133  \\ \hline
2                    & 0.01  & 0.10  & 750         & 0.7861  & 0.8079   & 0.8138   & 0.8136  \\ \hline
3                    & 0.01  & 0.15  & 1000        & 0.7869  & 0.8075   & 0.8152   & 0.8139  \\ \hline
4                    & 0.05  & 0.01  & 500         & 0.7869  & 0.8088   & 0.8156   & 0.8146  \\ \hline
5                    & 0.05  & 0.05  & 250         & 0.7869  & 0.8062   & 0.8144   & 0.8136  \\ \hline
6                    & 0.05  & 0.10  & 1000        & 0.7869  & 0.8088   & 0.8167   & 0.8148  \\ \hline
7                    & 0.05  & 0.15  & 750         & 0.7869  & 0.8088   & 0.8164   & 0.8147  \\ \hline
8                    & 0.10  & 0.01  & 750         & 0.7869  & 0.8088   & 0.8161   & 0.8148  \\ \hline
9                    & 0.10  & 0.05  & 1000        & 0.7869  & 0.8088   & 0.8170   & 0.8148  \\ \hline
10                   & 0.10  & 0.10  & 250         & 0.7869  & 0.8088   & 0.8161   & 0.8141  \\ \hline
11                   & 0.10  & 0.15  & 500         & 0.7869  & 0.8088   & 0.8170   & 0.8143  \\ \hline
12                   & 0.15  & 0.01  & 1000        & 0.7869  & 0.8088   & 0.8170   & 0.8148  \\ \hline
13                   & 0.15  & 0.05  & 750         & 0.7869  & 0.8088   & 0.8170   & 0.8148  \\ \hline
14                   & 0.15  & 0.10  & 500         & 0.7869  & 0.8088   & 0.8170   & 0.8148  \\ \hline
15                   & 0.15  & 0.15  & 250         & 0.7869  & 0.8088   & 0.8167   & 0.8141  \\ \hline
\end{tabular}%
}
\end{table}
\begin{figure}[tbp]
  \centering
  \subfigure[Mean for $L=8$]{\includegraphics[width=0.23\textwidth]{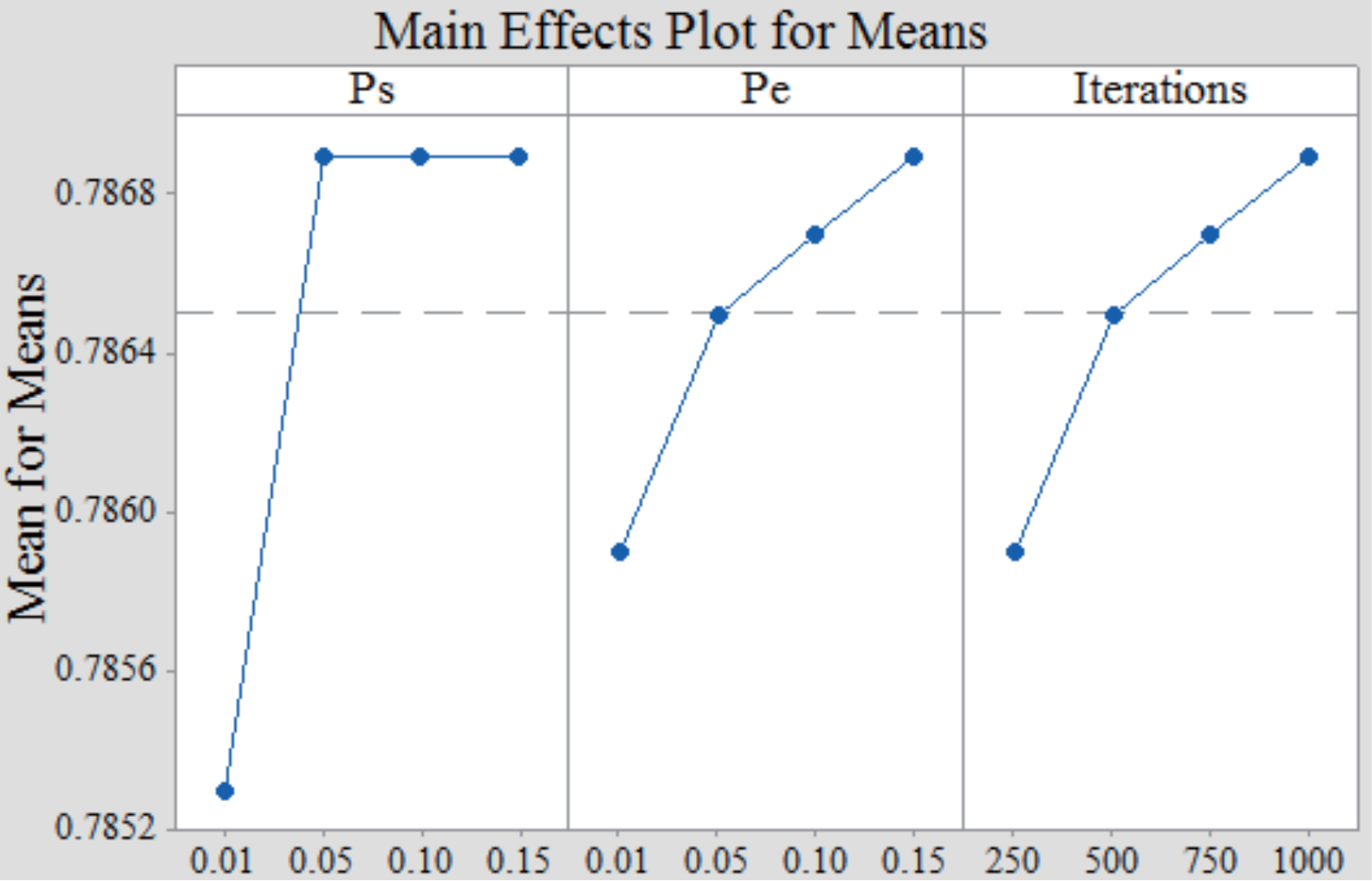}
  \label{Mean for 8 tasks}}
  \subfigure[Mean for $L=16$]{\includegraphics[width=0.23\textwidth]{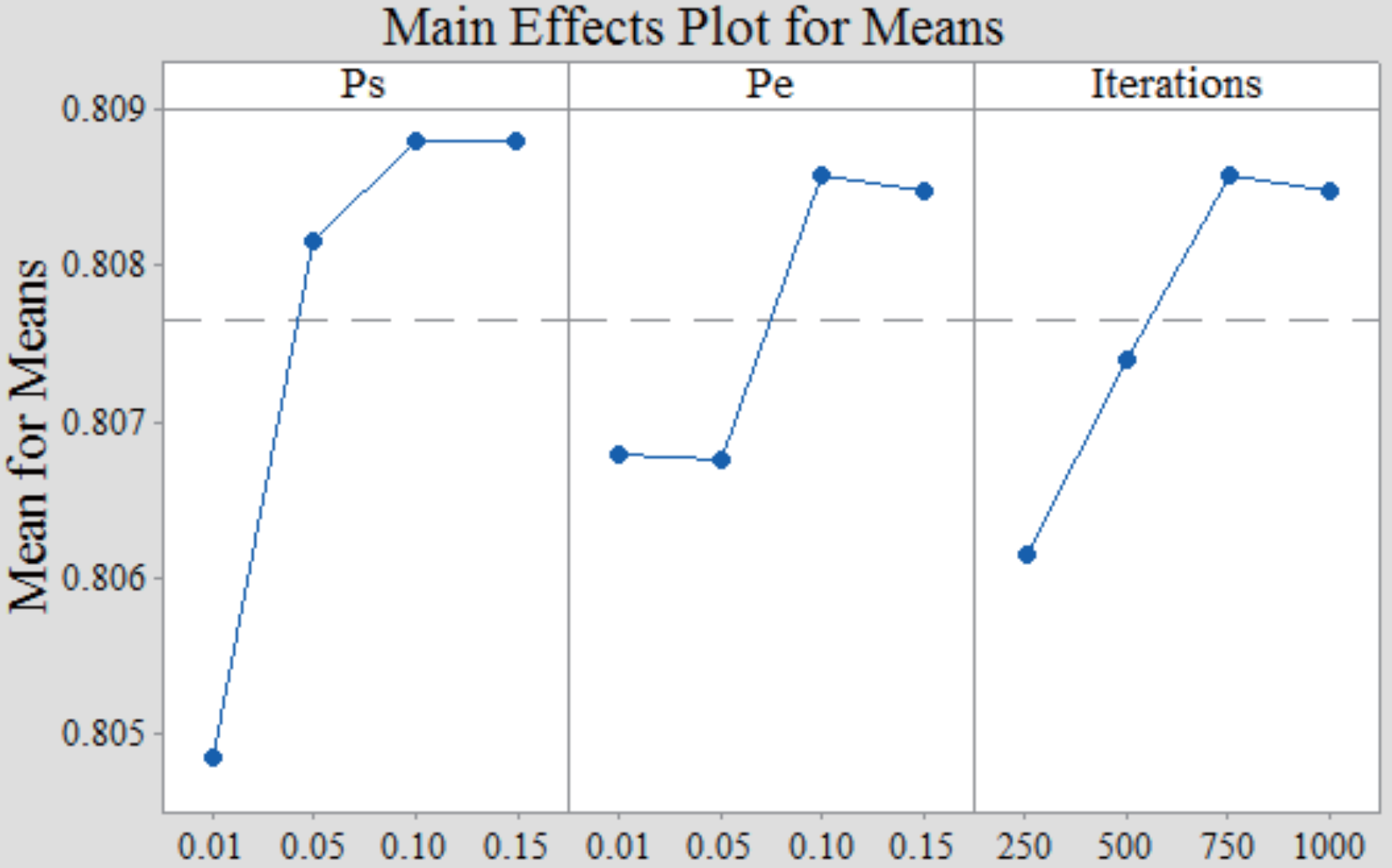}
  \label{Mean for 16 tasks}}
  \subfigure[Mean for $L=24$]{\includegraphics[width=0.23\textwidth]{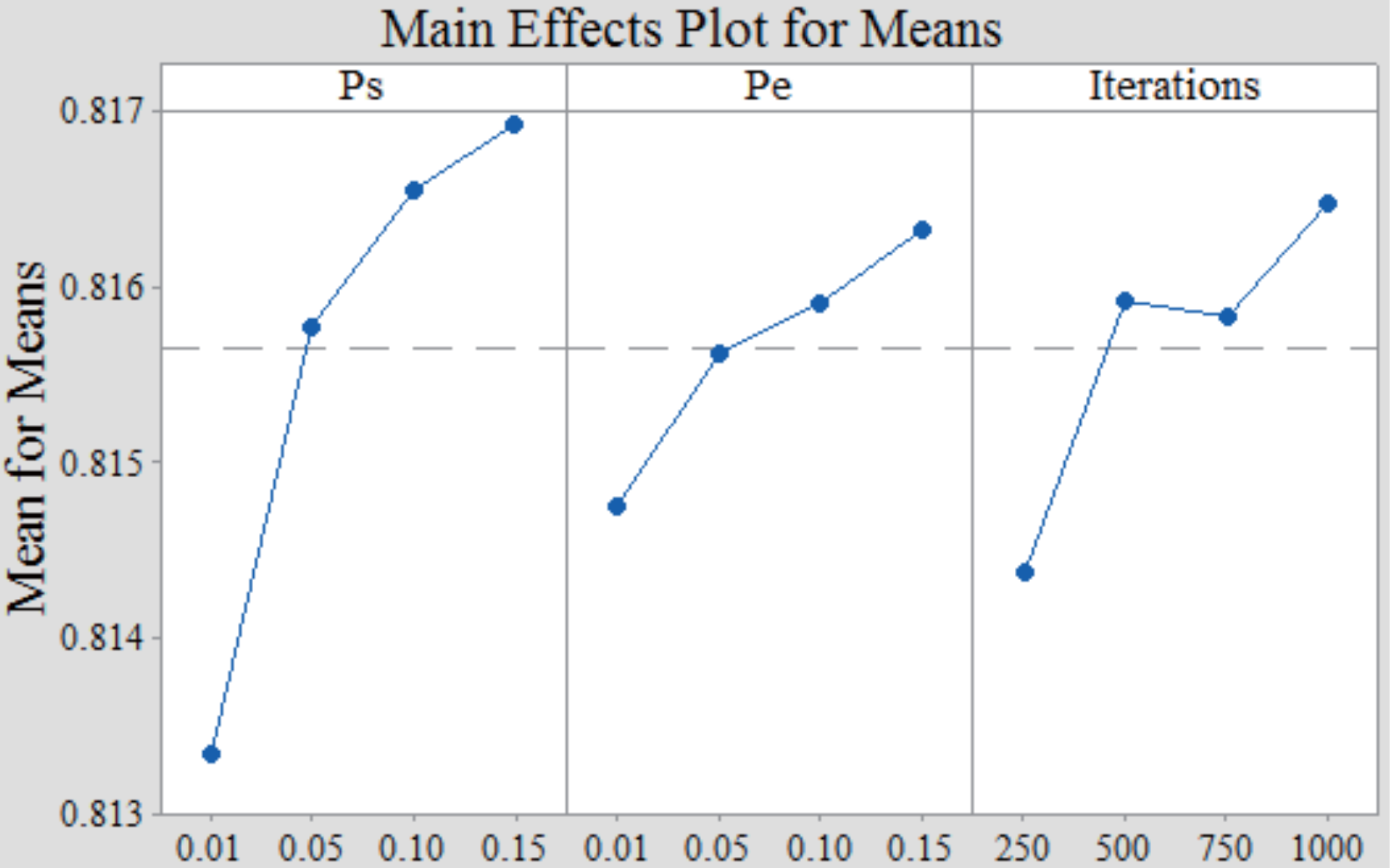}
  \label{Mean for 4 tasks}}
  \subfigure[Mean for $L=32$]{\includegraphics[width=0.23\textwidth]{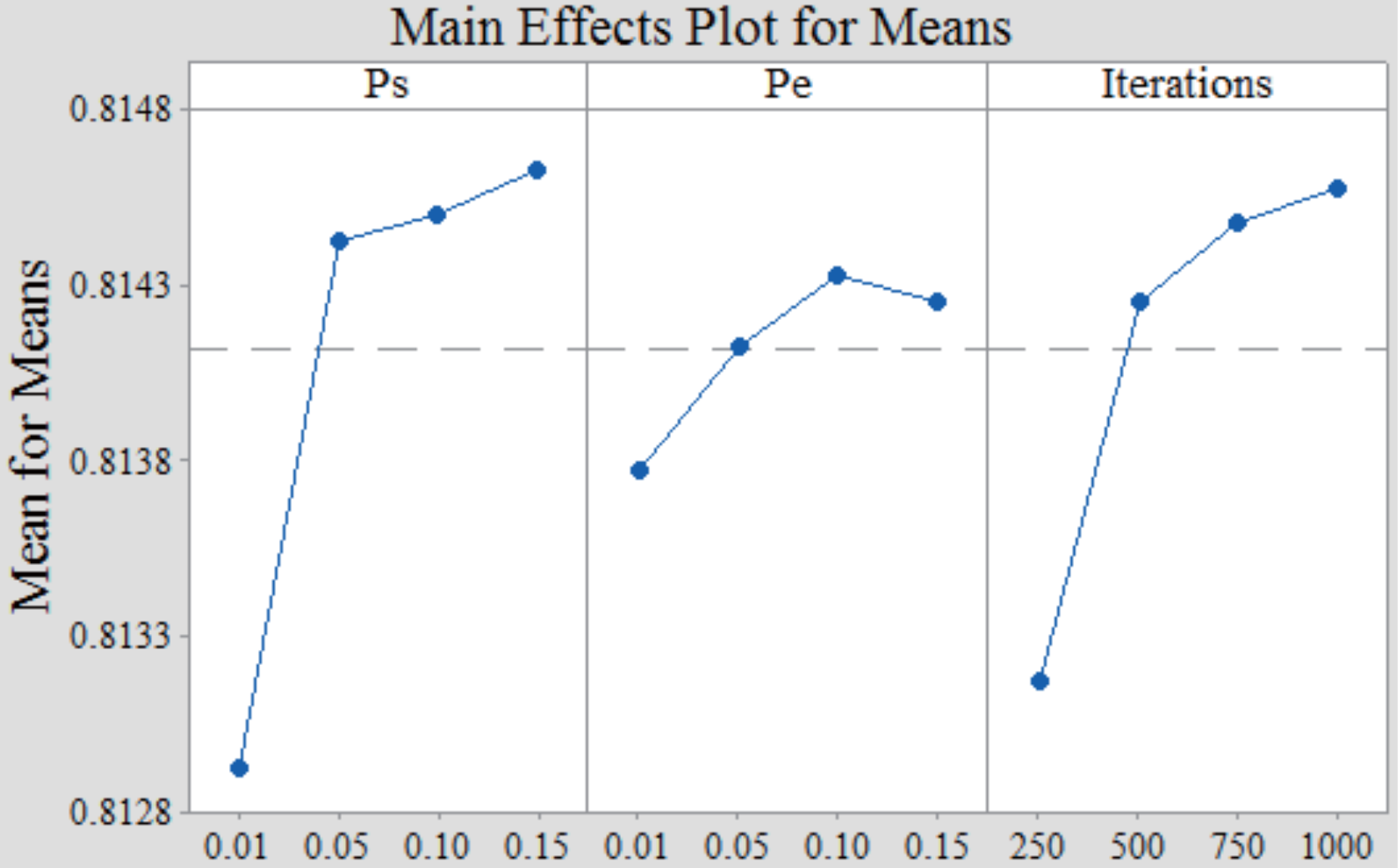}
  \label{Mean for 32 tasks}}
  \caption{Mean for $L=8,16,24,32$.}
  \label{Mean for different tasks}
\end{figure}

Fig. \ref{Mean for different tasks} shows the main effects plot for means with different tasks. It can be obvious that the objective values are improved as $p_s,p_e$ and the number of iterations increase. For the proposed HMAO algorithm, we can adjust the values of $p_s,p_e$ to the diversity of feasible solutions, however, it will cost more time to seek an approximate solution for $Z(X)$. The ideal value of $p_s$ for 8, 16, 24, 32 tasks is 0.15, the ideal value of $p_e$ is 0.1 for 8, 24 tasks and 0.15 for 16, 32 tasks.

\subsection{Numerical Simulation}\label{Numerical Simulation}

\begin{figure}[tbp]
  \centering
  \subfigure[$L=8,p_s=0.15,p_e=0.15$]{\includegraphics[width=0.23\textwidth]{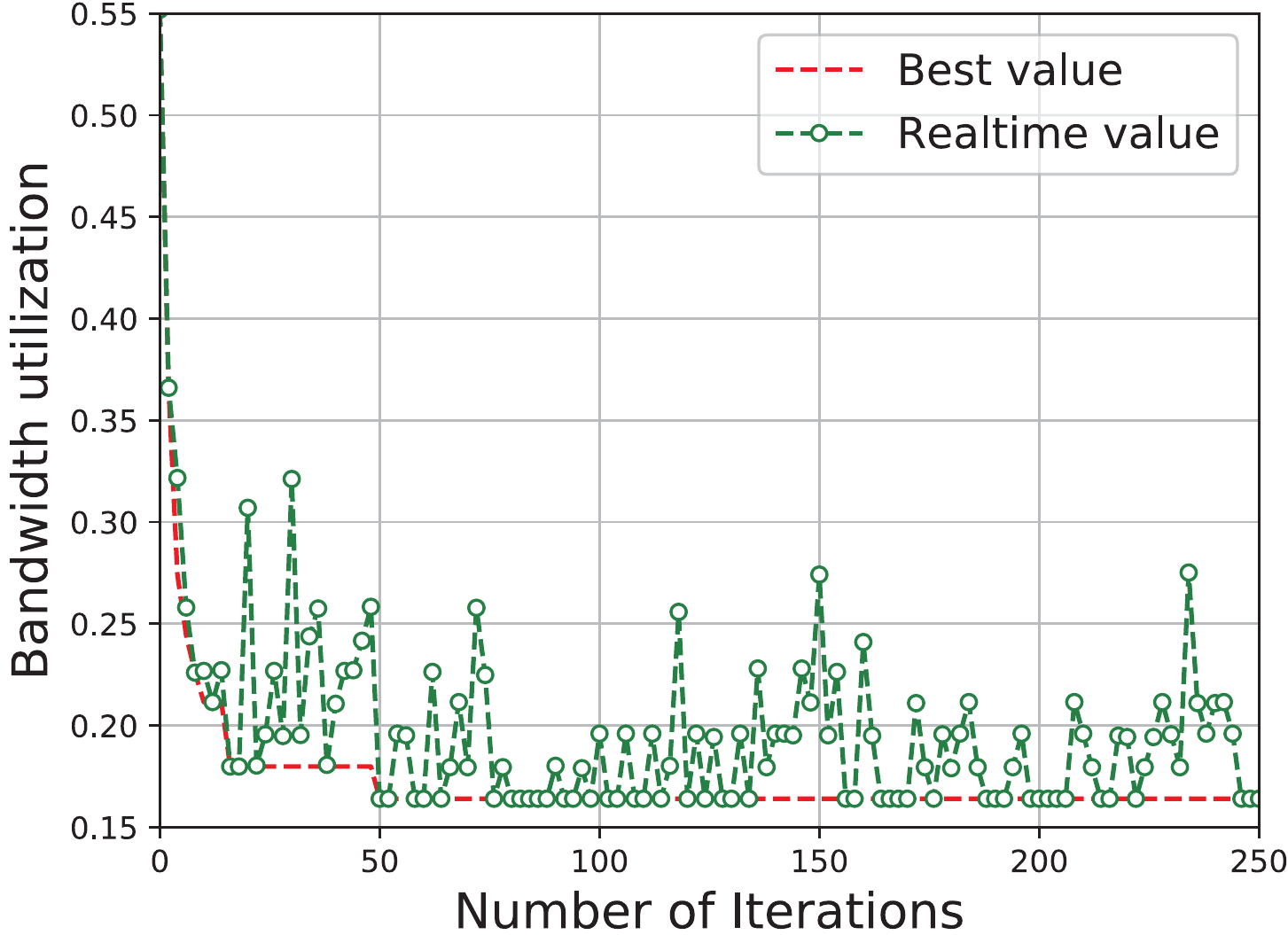}
  \label{Bandwidth utilization for 8 tasks}}
  \subfigure[$L=16,p_s=0.15,p_e=0.10$]{\includegraphics[width=0.23\textwidth]{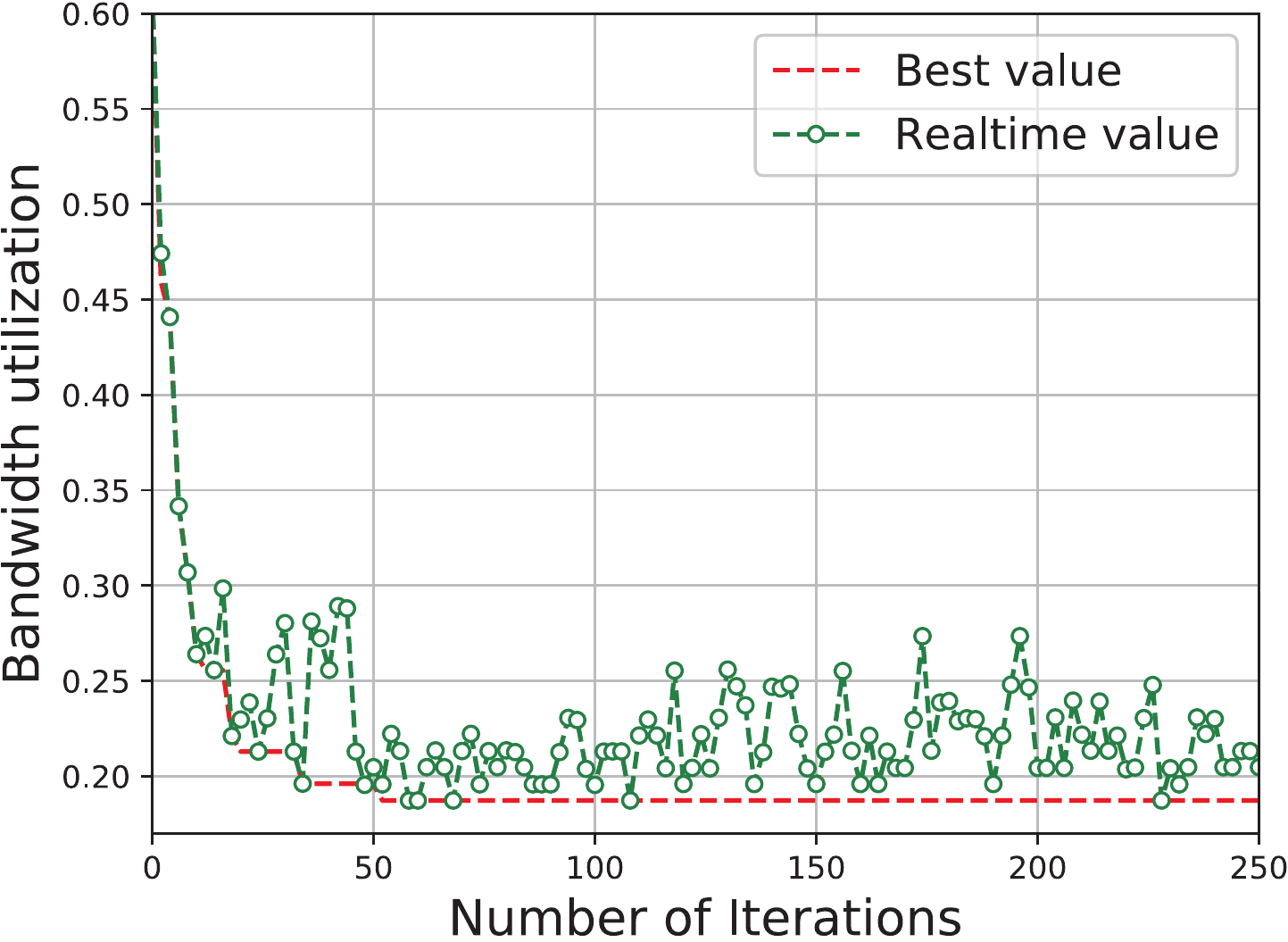}
  \label{Bandwidth utilization for 16 tasks}}
  \subfigure[$L=24,p_s=0.15,p_e=0.15$]{\includegraphics[width=0.23\textwidth]{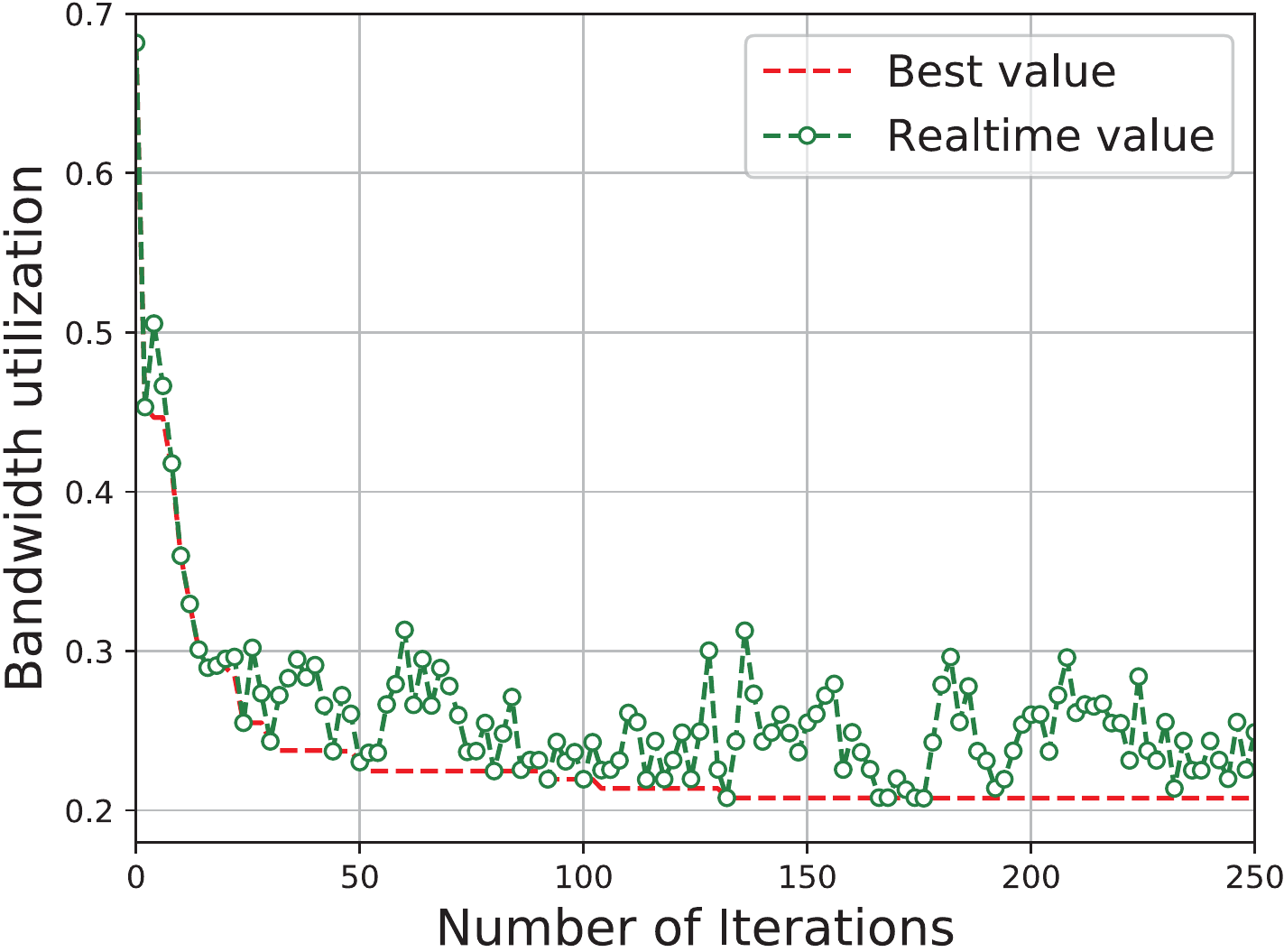}
  \label{Bandwidth utilization for 24 tasks}}
  \subfigure[$L=32,p_s=0.15,p_e=0.10$]{\includegraphics[width=0.23\textwidth]{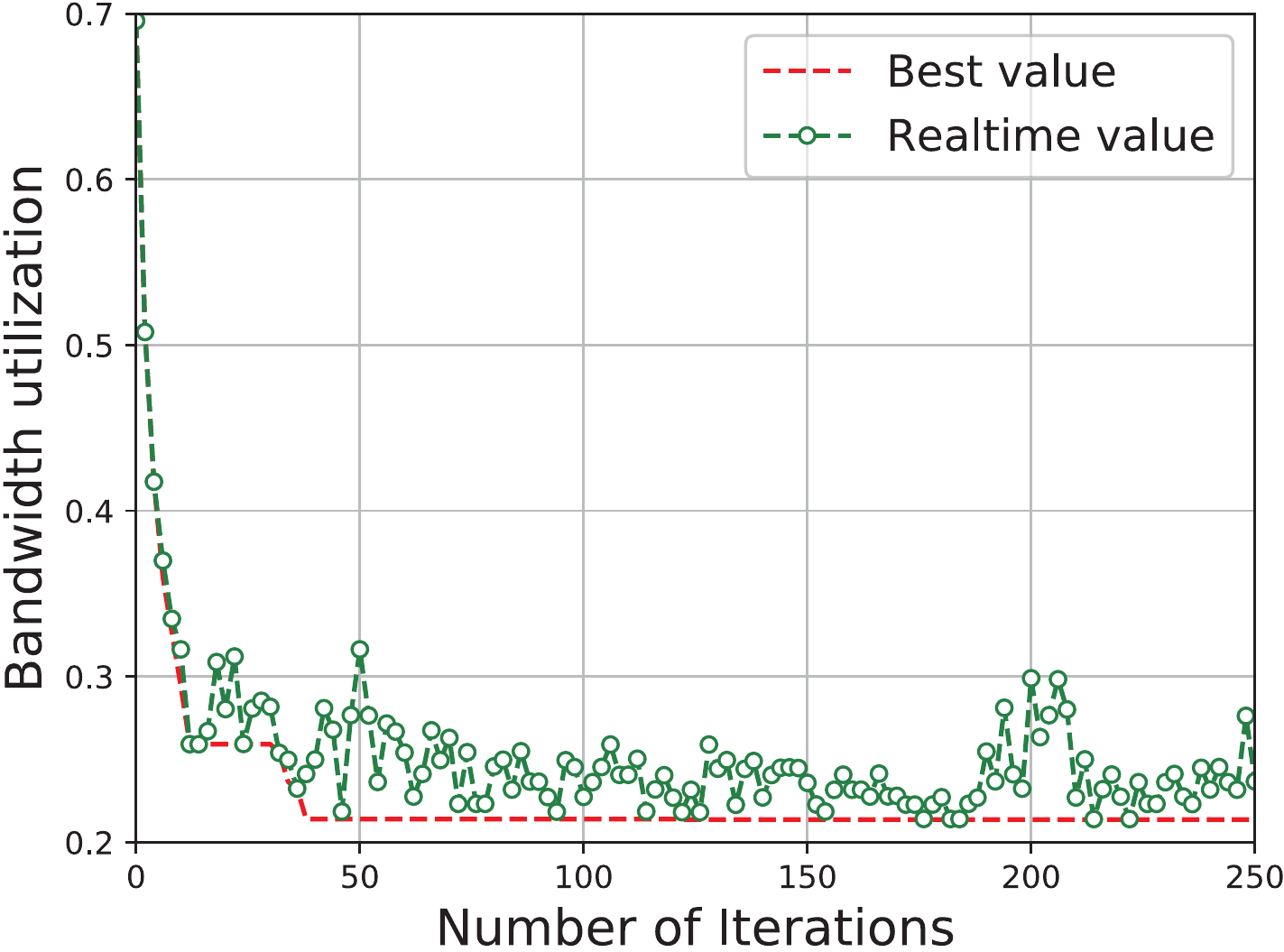}
  \label{Bandwidth utilization for 32 tasks}}
  \caption{Bandwidth utilization for $L=8,16,24,32$.}
  \label{Bandwidth utilization for 8,16,24,32 tasks}
\end{figure}

Based on the parameters above, the maximum number of iterations is set as $250$, we run several experiments for 8, 16, 24, 32 tasks to evaluate convergence of the proposed HMAO algorithm. As it is shown in Fig. \ref{Bandwidth utilization for 8,16,24,32 tasks}, we illustrate the relationships between iterations and the bandwidth utilization for different tasks. Fig. \ref{Bandwidth utilization for 8 tasks} and Fig. \ref{Bandwidth utilization for 24 tasks} indicate the procedure for seeking the optimal solution of bandwidth utilization for $L=8, 24$ with $p_s=0.15$, $p_e=0.15$, and the best values 0.1640 and 0.0.2076 can be observed at 49 and 176 iterations, respectively. For $L=16,32$, $p_s=0.15$, $p_e=0.10$, their results for optimizing the bandwidth utilization are depicted in Fig. \ref{Bandwidth utilization for 16 tasks} and Fig. \ref{Bandwidth utilization for 32 tasks}, and we can obtain the optimal solution results 0.1872 and 0.2137 at 73 and 121 iterations, respectively.\par
\begin{table}[tbp]
\centering
\caption{Results of simulation for the HMAO algorithm.}
\label{Results of simulation for the HMAO algorithm}
\resizebox{\columnwidth}{!}{%
\begin{tabular}{|c|c|c|c|c|c|c|}
\hline
$L$ & $M$           & Min    & Max    & Mean   & Std    & Time(min) \\ \hline
8    & \multirow{4}{*}{250} & 0.7869 & 0.7869 & 0.7869 & 0.0    & 0.2375    \\ \cline{1-1} \cline{3-7}
16   &                      & 0.8046 & 0.8088 & 0.8084 & 0.0013 & 0.4777    \\ \cline{1-1} \cline{3-7}
24   &                      & 0.8141 & 0.8170 & 0.8164 & 0.0012 & 0.7205    \\ \cline{1-1} \cline{3-7}
32   &                      & 0.8104 & 0.8148 & 0.8137 & 0.0015 & 0.9703    \\ \hline
8    & \multirow{4}{*}{500} & 0.7869 & 0.7869 & 0.7869 & 0.0    & 0.4995    \\ \cline{1-1} \cline{3-7}
16   &                      & 0.8088 & 0.8088 & 0.8088 & 0.0    & 1.0125    \\ \cline{1-1} \cline{3-7}
24   &                      & 0.8170 & 0.8170 & 0.8170 & 0.0    & 1.5177    \\ \cline{1-1} \cline{3-7}
32   &                      & 0.8146 & 0.8148 & 0.8147 & 0.0001 & 2.0164    \\ \hline
\end{tabular}%
}
\end{table}

\begin{figure}[tbp]
  \centering
  \subfigure[$L=8,p_s=0.15,p_e=0.15$]{\includegraphics[width=0.23\textwidth]{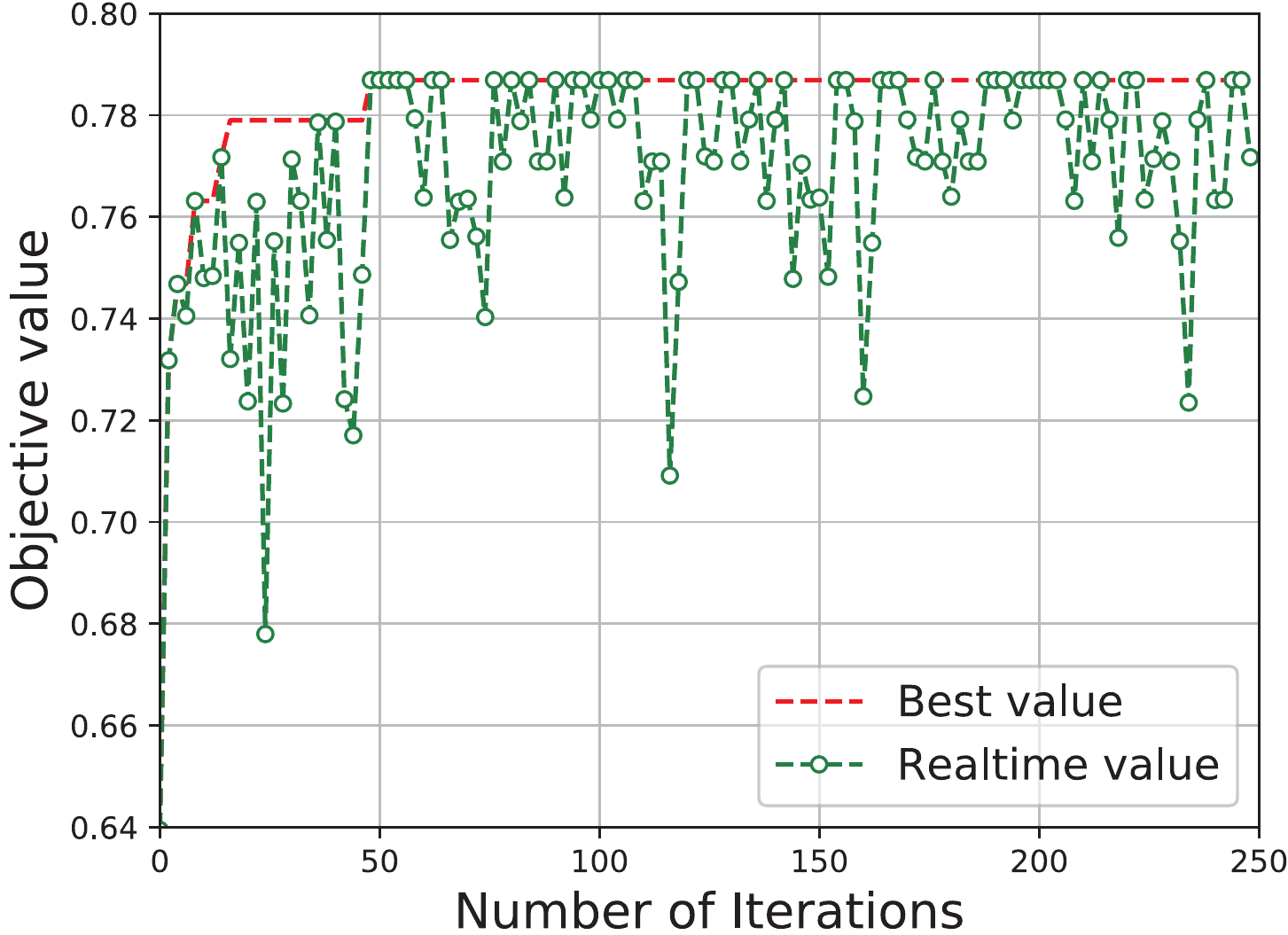}
  \label{Objective values for 8 tasks}}
  \subfigure[$L=16,p_s=0.15,p_e=0.10$]{\includegraphics[width=0.23\textwidth]{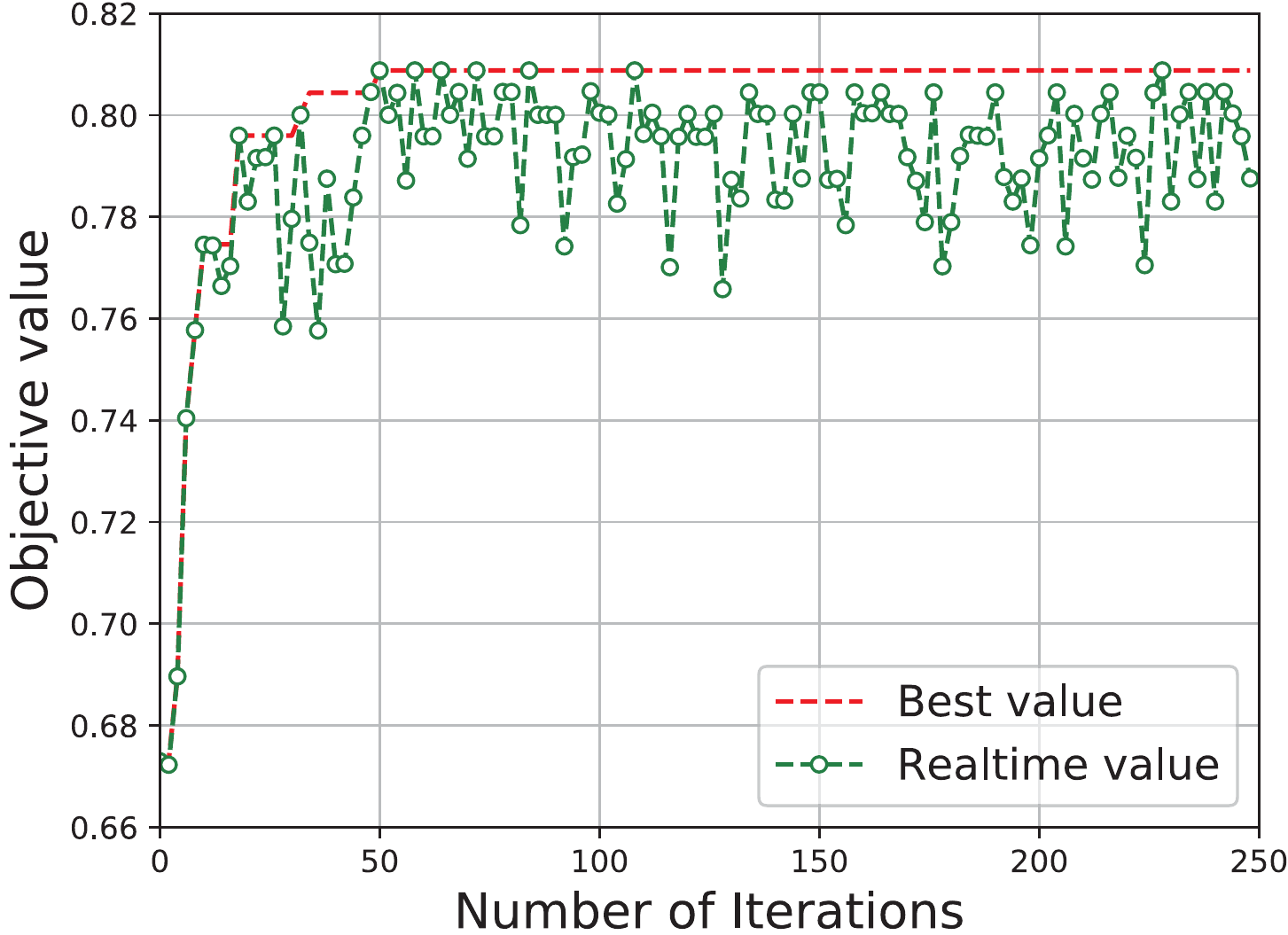}
  \label{Objective values for 16 tasks}}
  \subfigure[$L=24,p_s=0.15,p_e=0.15$]{\includegraphics[width=0.23\textwidth]{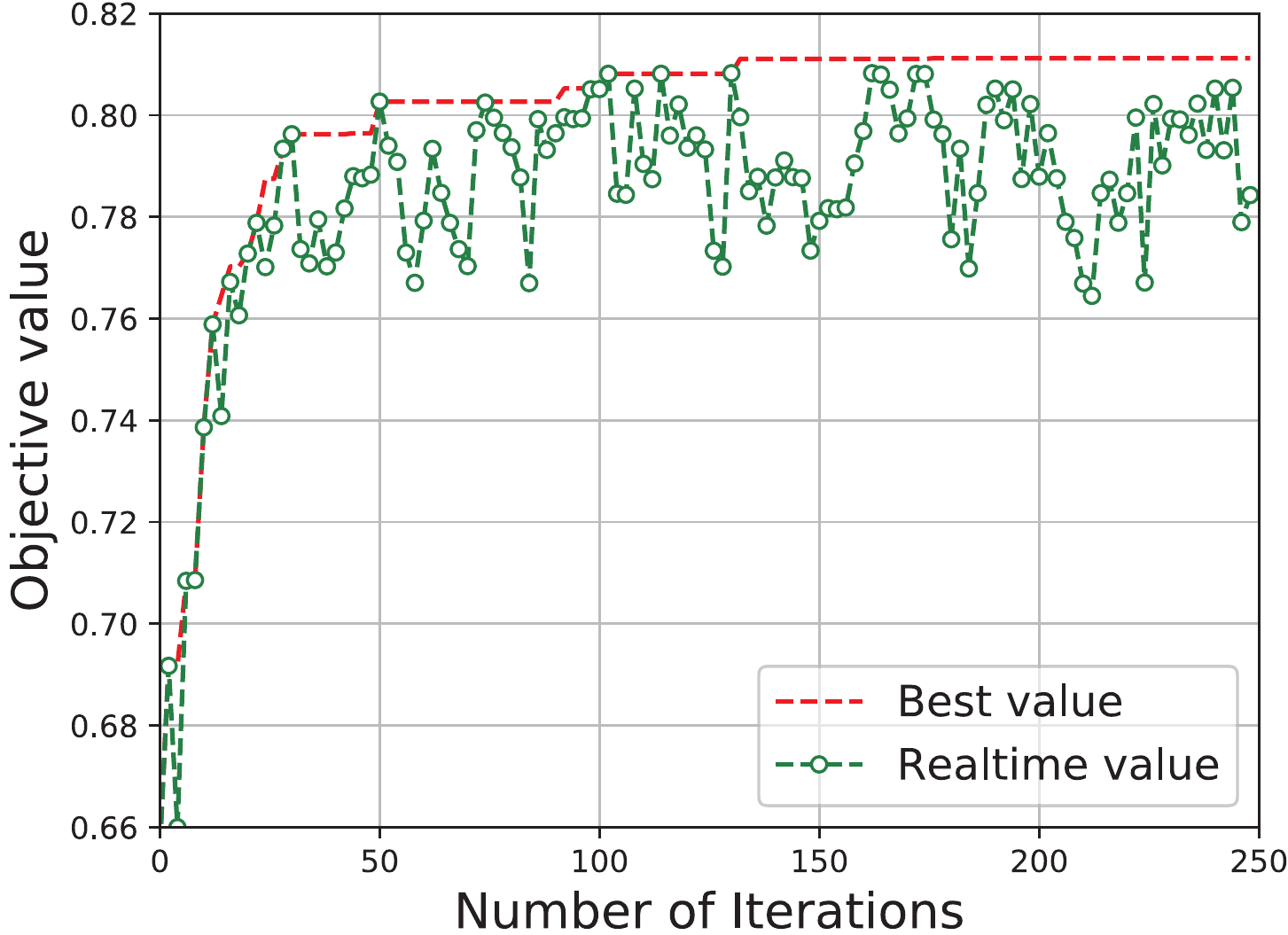}
  \label{Objective values for 24 tasks}}
  \subfigure[$L=32,p_s=0.15,p_e=0.10$]{\includegraphics[width=0.23\textwidth]{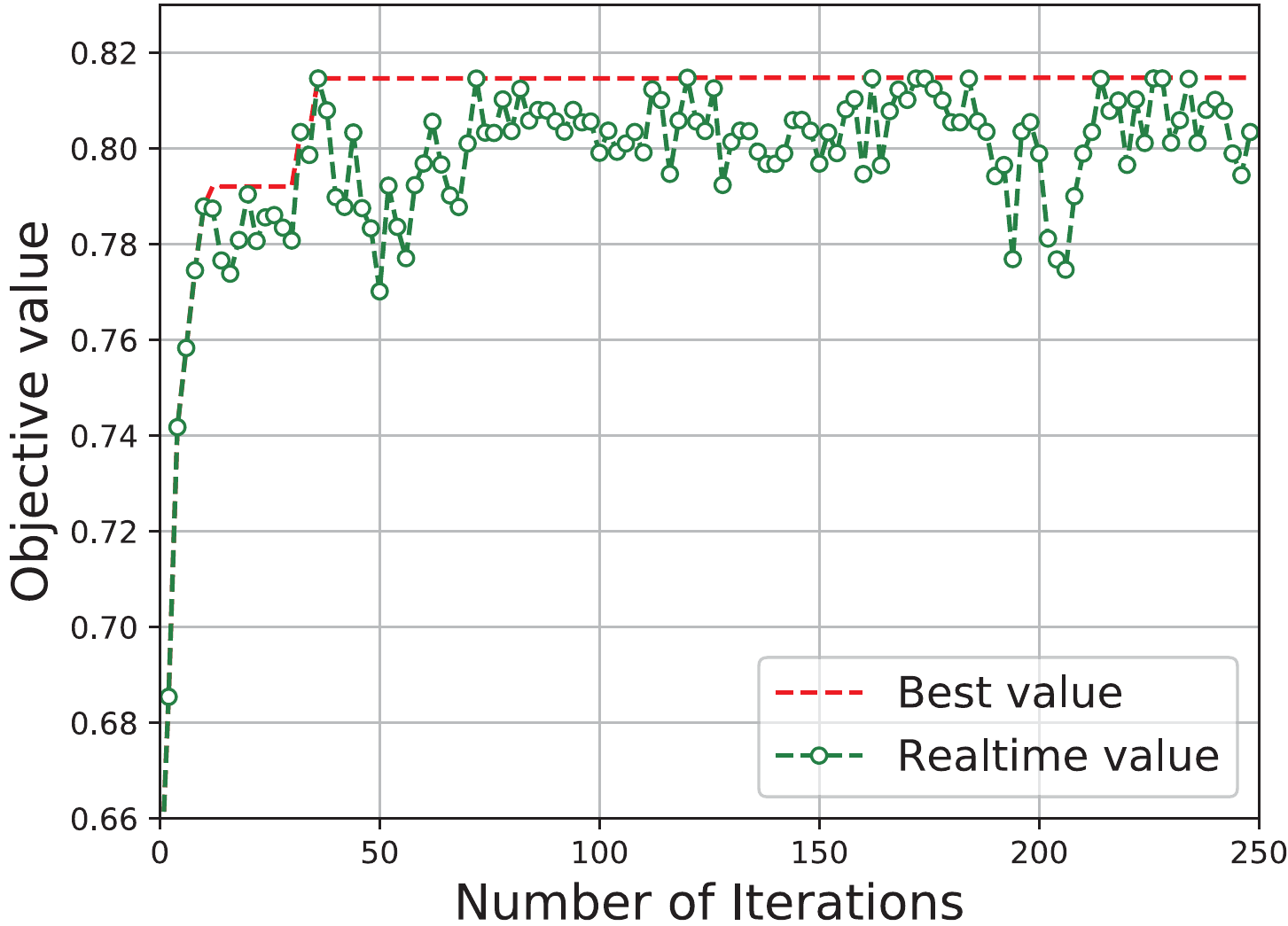}
  \label{Objective values for 32 tasks}}
  \caption{Objective values for $L=8,16,24,32$.}
  \label{Objective values for 8,16,24,32 tasks}
\end{figure}

Similarly, Fig. \ref{Objective values for 8,16,24,32 tasks} describes the evolutionary plots of the iterative optimal solutions for different tasks, which include the best and real-time objective values. Fig. \ref{Objective values for 8 tasks} and Fig. \ref{Objective values for 24 tasks} show the simulation results of 8, 24 tasks with $p_s=0.15$, $p_e=0.15$, Fig. \ref{Objective values for 16 tasks} and Fig. \ref{Objective values for 32 tasks} show the simulation results of 16, 32 tasks with $p_s=0.15$, $p_e=0.10$. It can be observed that the objective values for 8, 16, 24 and 32 tasks converge to the approximate solution results as 0.7869, 0.8088, 0.8112 and 0.8148 at 49, 73, 176 and 121 iterations, respectively. Therefore, it can be depicted that the proposed HMAO algorithm is an effective optimization algorithm to solve the problem of resource allocation in cloud computing and has a better convergence performance.\par

As the number of iterations have impact on the performance of the proposed HMAO algorithm, therefore, some experiments with 250 and 500 iterations are carried out for different tasks. Each case runs 10 times, we can obtain minimum, maximum, mean and standard deviation of the optimal solutions and the average computing time, which are shown in Table \ref{Results of simulation for the HMAO algorithm}. It can be observed from Table \ref{Results of simulation for the HMAO algorithm} that the performance of the proposed HMAO algorithm is high when the number of iterations increases, i.e., the means for 8, 16, 24 and 32 tasks with 250 iterations are 0.7869, 0.8084, 0.8164 and 0.8137, respectively, while the means are 0.7869, 0.8088, 0.8170 and 0.8147 with 500 iterations, respectively. That is, the potential optimal solution for $Z(X)$ is more likely to be sought by exploring and exploiting the solution space iteratively. Besides, the results indicate that the computing time of the proposed HMAO algorithm increases almost linearly with the increase in the number of iterations.\par

\subsection{Performance Comparison with the Baseline Algorithms}\label{Performance Comparison with the Baseline Algorithms}

In order to further discuss the effectiveness of the proposed HMAO algorithm, we compare the proposed HMAO algorithm for different tasks with two existing algorithms, which are GA and NSGA-II.\par

\textbf{GA:} The genetic algorithm described in \cite{tseng2017dynamic} is used in our paper. We randomly select two individuals from the population $P$ to be the parents, and they can mate with each other by the two-point crossover operator with crossover probability $p_c$ to generate their offsprings. If the fitness value of an offspring is superior to its parent, we will consider it as a candidate in the next generation. Otherwise, the mutation operator is carried out with mutation probability $p_m$. An individual that has a better fitness value between the offspring and its parent will be seen as a new individual in the next generation.\par

\textbf{NSGA-II:} We introduce NSGA-II in \cite{deb2002fast} to address our problem, two optimization sub-problems are $Z_{1}(X)$ and $1-Z_{2}(X)$, respectively. A binary tournament selection method \cite{razali2011genetic} is applied to make decision for choosing the parents from the population $P$ to mate with each other. In crossover operator, we randomly select two gene points of one individual to exchange their genes equivalently with the other individual under crossover probability $p_c$. A bitwise mutation is executed with mutation probability $p_m$.\par

\begin{figure}[tbp]
  \centering
  \subfigure[$L=4$]{\includegraphics[width=0.23\textwidth]{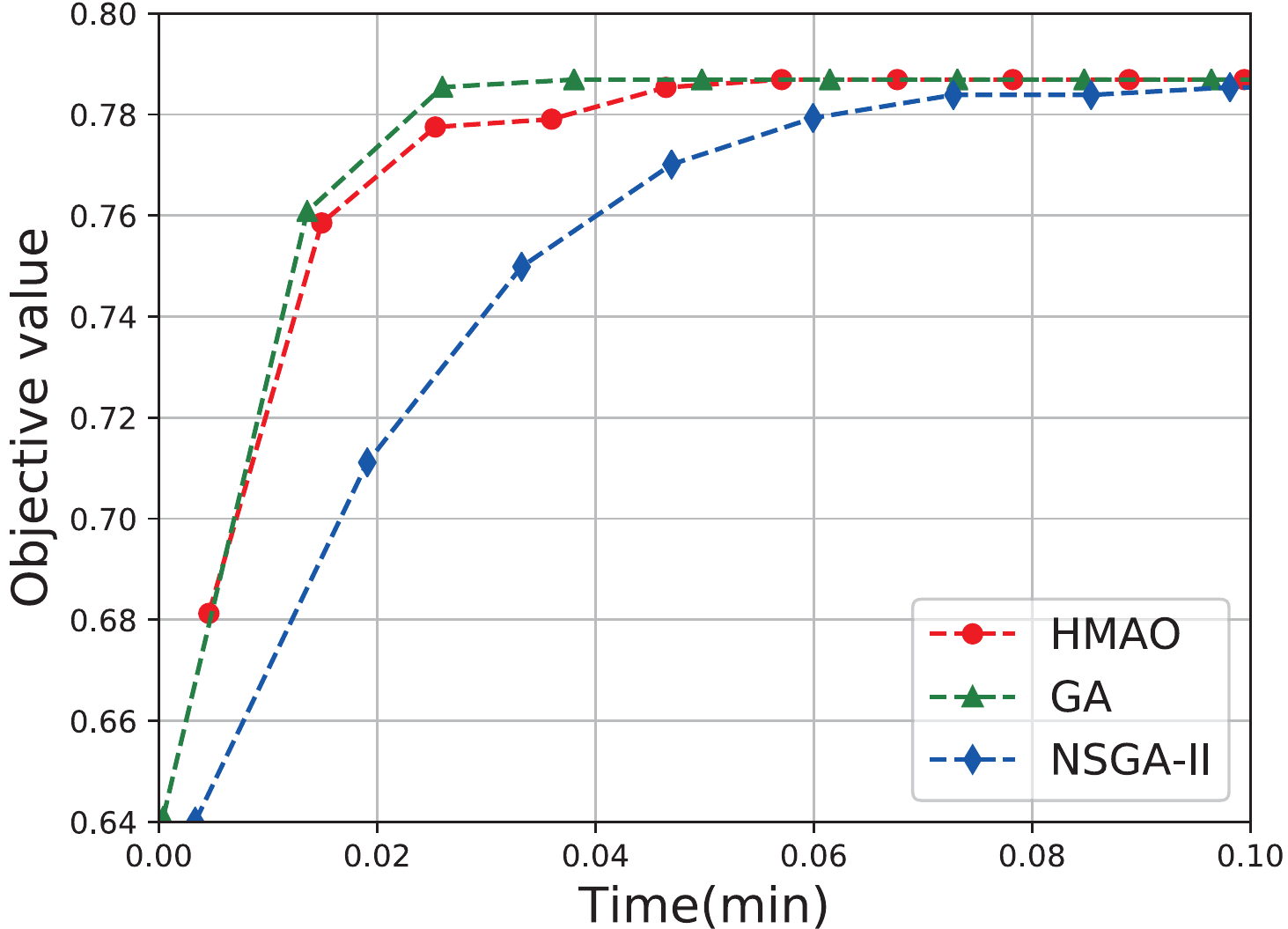}
  \label{Evolution of optimal solution for 4 tasks with different algorithms}}
  \subfigure[$L=6$]{\includegraphics[width=0.23\textwidth]{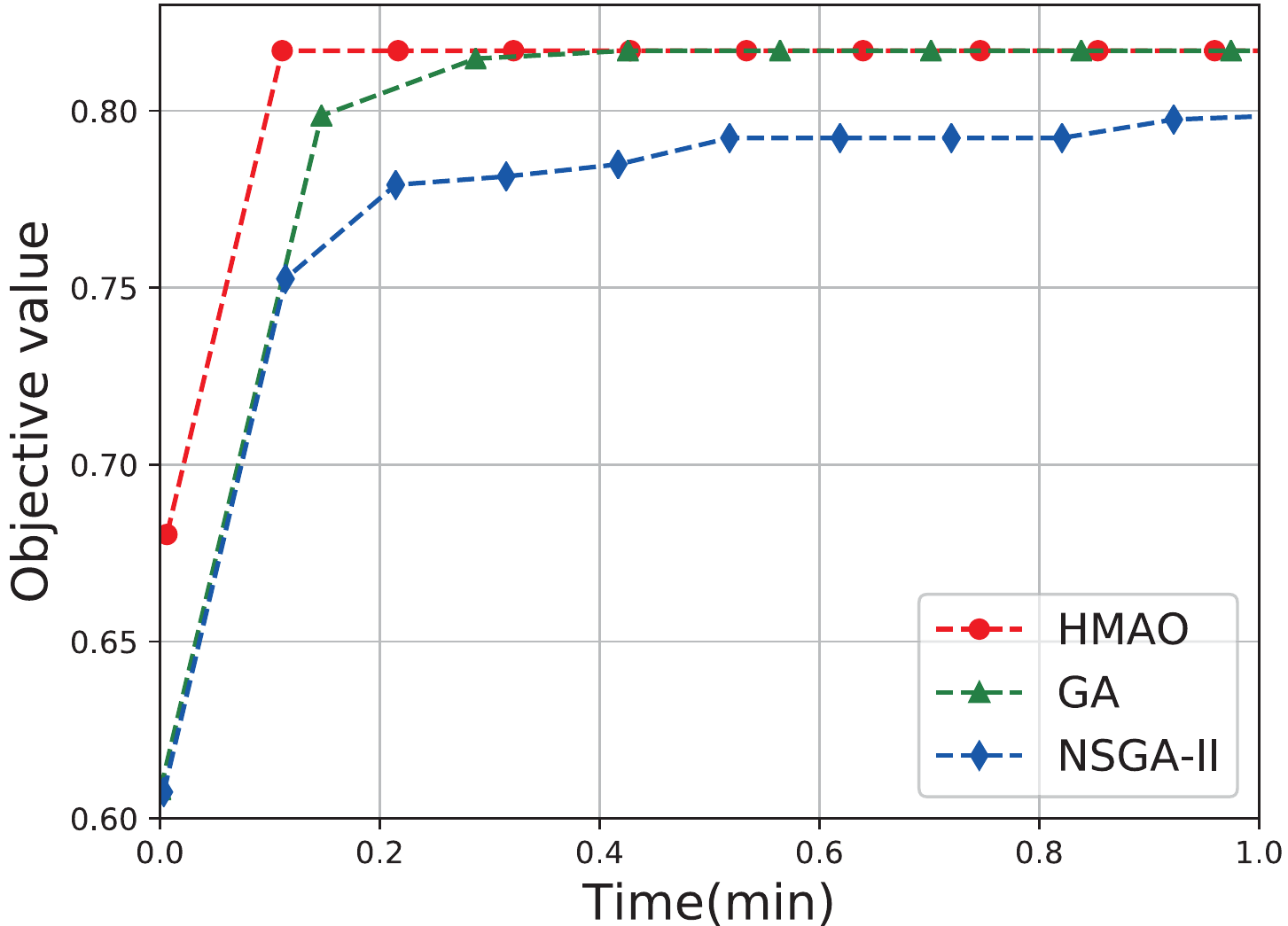}
  \label{Evolution of optimal solution for 6 tasks with different algorithms}}
  \subfigure[$L=8$]{\includegraphics[width=0.23\textwidth]{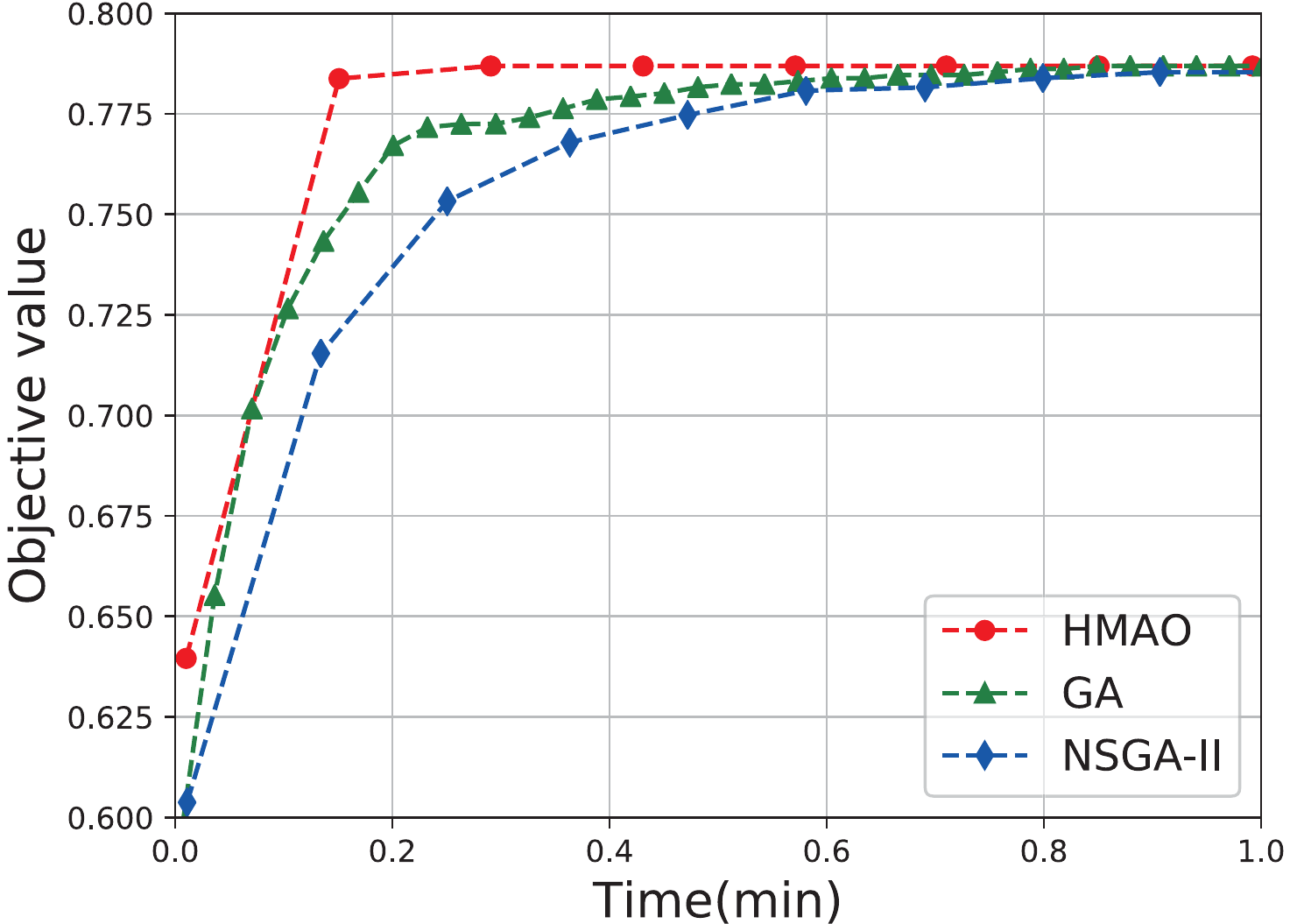}
  \label{Evolution of optimal solution for 8 tasks with different algorithms}}
  \subfigure[$L=10$]{\includegraphics[width=0.23\textwidth]{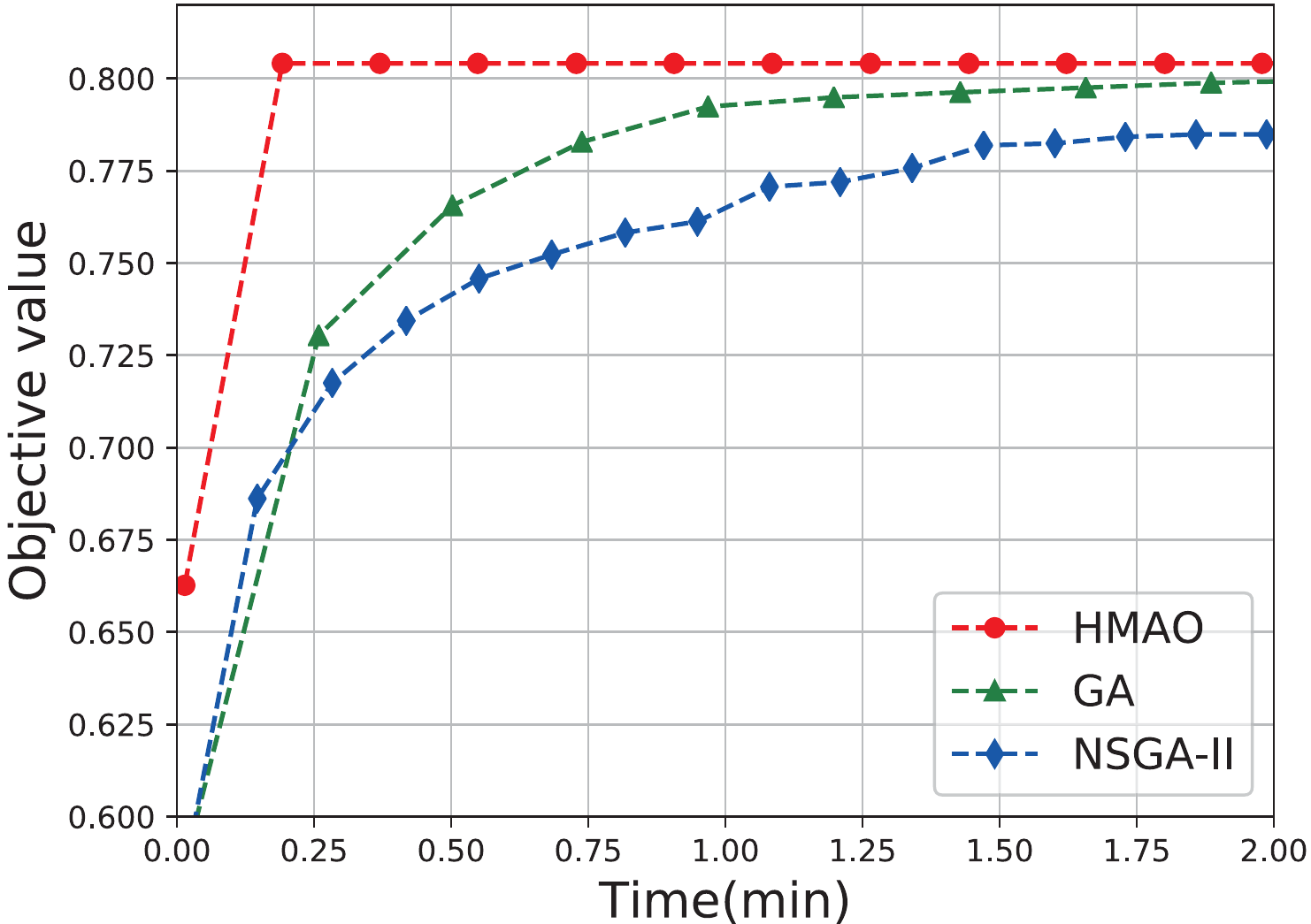}
  \label{Evolution of optimal solution for 10 tasks}}
  \caption{Objective values for $L=4,6,8,10$ with HMAO, GA and NSGA-II.}
  \label{Evolution of optimal solution}
\end{figure}

With the above analysis of the proposed HMAO algorithm, we set the maximum iterations as 1000 for the proposed HMAO algorithm, 5000 for GA and NSGA-II, the population $P$ is set as 16 for the proposed HMAO algorithm, 100 for GA and NSGA-II. Moreover, we have $p_e=p_s=0.15$, $p_c=1.0$, $p_m=0.1$. The number of tasks is from 4 to 16 corresponding to the number of initial service nodes $4,5,\dots,16$, respectively. We run each test case 10 times and compute the average results. The parameters used for the proposed HMAO algorithm, GA and NSGA-II are summarized in Table \ref{Parameter setting for HMAO, GA and NSGA-II}.\par

\begin{table}[tbp]
\centering
\caption{Parameter setting for HMAO, GA and NSGA-II.}
\label{Parameter setting for HMAO, GA and NSGA-II}
\resizebox{\columnwidth}{!}{%
\begin{tabular}{|c|c|}
\hline
Parameters                      & Value                                                 \\ \hline
Number of tasks $L$             & $\left \{ 4,5,6,7,8,9,10,11,12,13,14,15,16 \right \}$   \\ \hline
Number of service nodes $K$     & $\left \{ 4,5,6,7,8,9,10,11,12,13,14,15,16 \right \}$   \\ \hline
Maximum iterations $M$           & HMAO: 1000; GA,NSGA-II: 5000                    \\ \hline
Population size $P$             & HMAO: 16; GA,NSGA-II: 100                       \\  \hline
Selective probability $p_e$     & 0.15                                                  \\ \hline
Exchange probability $p_s$      & 0.15                                                  \\ \hline
Crossover probability $p_c$     & 1.0                                                   \\ \hline
Mutation probability $p_m$      & 0.1                                                   \\ \hline
Running time                    & 10                                                    \\ \hline
\end{tabular}%
}
\end{table}
Firstly, we simulate all test cases by the proposed HMAO algorithm, GA and NSGA-II with the specified parameters in Table \ref{Parameter setting for HMAO, GA and NSGA-II} and obtain the average results of these three algorithms, including resource utilization, bandwidth utilization, objective value and time cost. Note that the number of initial service nodes $K$ can influence the results of GA and NSGA-II.\par

To investigate the evolution of the optimal solution for the proposed HMAO algorithm over time, we compare the evolutionary performance of the proposed HMAO algorithm, GA and NSGA-II for $L=4,6,8,10$ and the results of these three algorithms are shown in Fig. \ref{Evolution of optimal solution}. It can be observed that the performance of the proposed HMAO algorithm is similar to that of GA for $L=4,6,8$, NSGA-II for $L=4,8$. Hence, the effectiveness of the proposed HMAO algorithm is verified by comparing the performance of the proposed HMAO algorithm with GA and NSGA-II algorithms. Moreover, the convergence of the proposed HMAO algorithm is better than GA and NSGA-II as the number of tasks grows.\par

The reason is that the selection and exchange operators for the proposed HMAO algorithm are executed with a finer-grained sub-task level and the optimization objective tends to a good result with high probability during each iterative evolution. Besides, the probabilistic-based selection and exchange operators are used to achieve the diversity of feasible solutions and avoid early entering into the local optimal solution. However, the space of feasible solutions for GA and NSGA-II becomes larger with the increase in the number of tasks, i.e., it will take more time to search the optimal solution. In addition, Fig. \ref{Evolution of optimal solution} shows that GA outperforms NSGA-II for $L=4,6,8,10$, because each offspring generated by GA, which has a better fitness value than its parent, will be selected as a new individual in the next generation and that can guarantee that the performance of solving the optimization problem is improved in each iterative evolution.\par

\begin{figure}[tbp]
  \centering
  \subfigure[Resource utilization]{\includegraphics[width=0.23\textwidth]{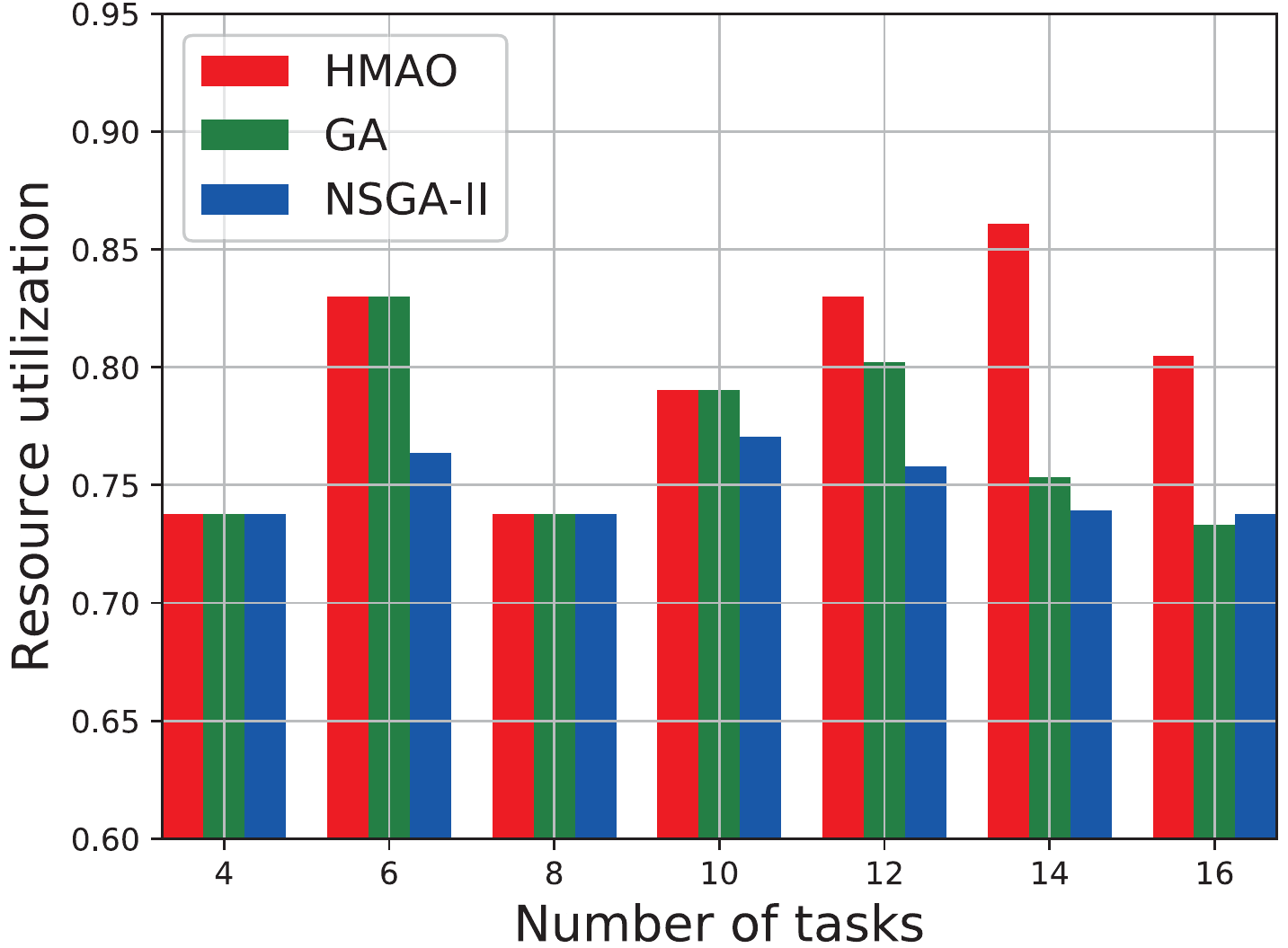}
  \label{Resource utilization}}
  \subfigure[Bandwidth utilization]{\includegraphics[width=0.23\textwidth]{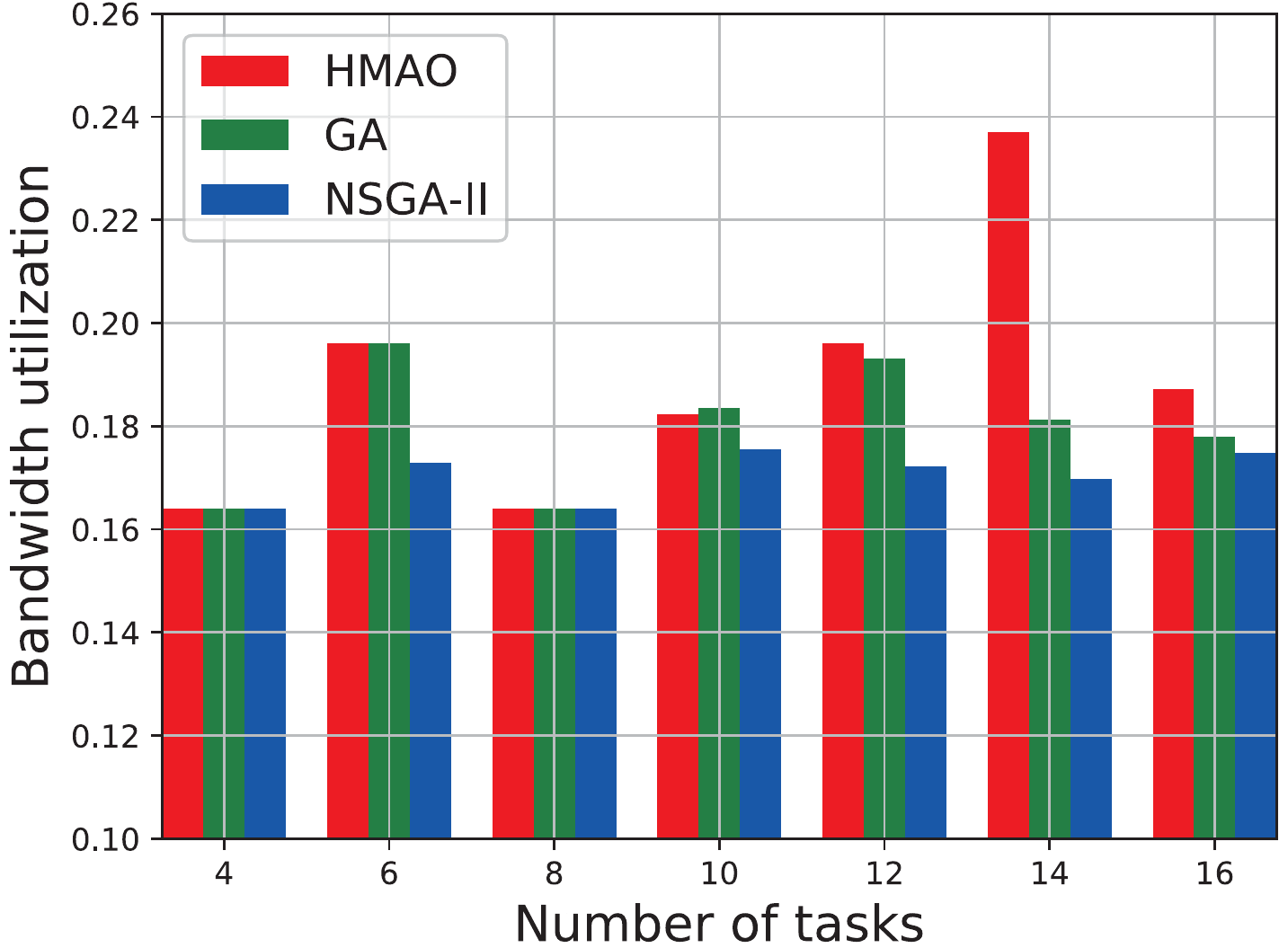}
  \label{Bandwidth utilization}}
  \subfigure[Optimal solution]{\includegraphics[width=0.23\textwidth]{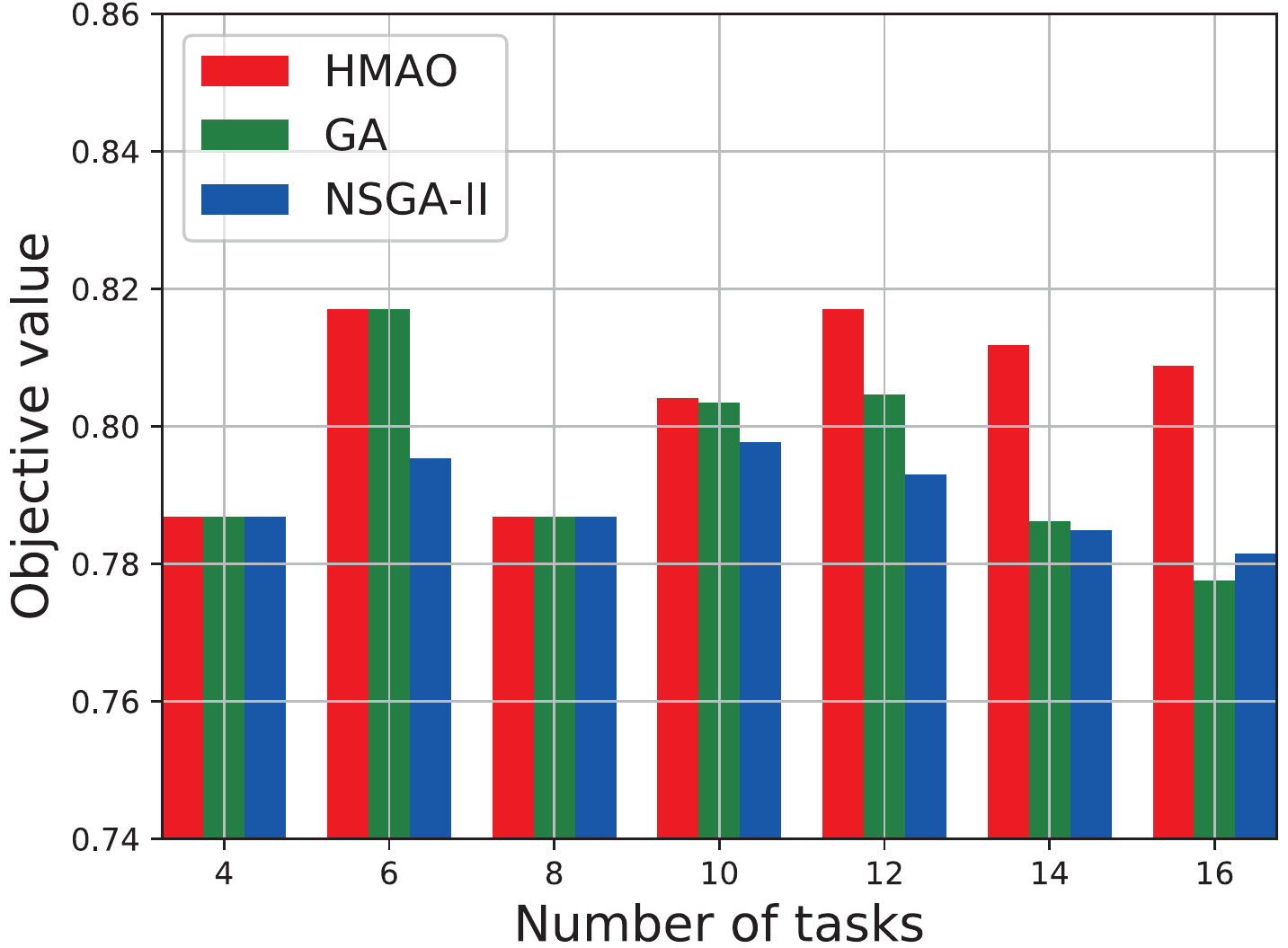}
  \label{Optimal solution}}
  \subfigure[Convergence time]{\includegraphics[width=0.23\textwidth]{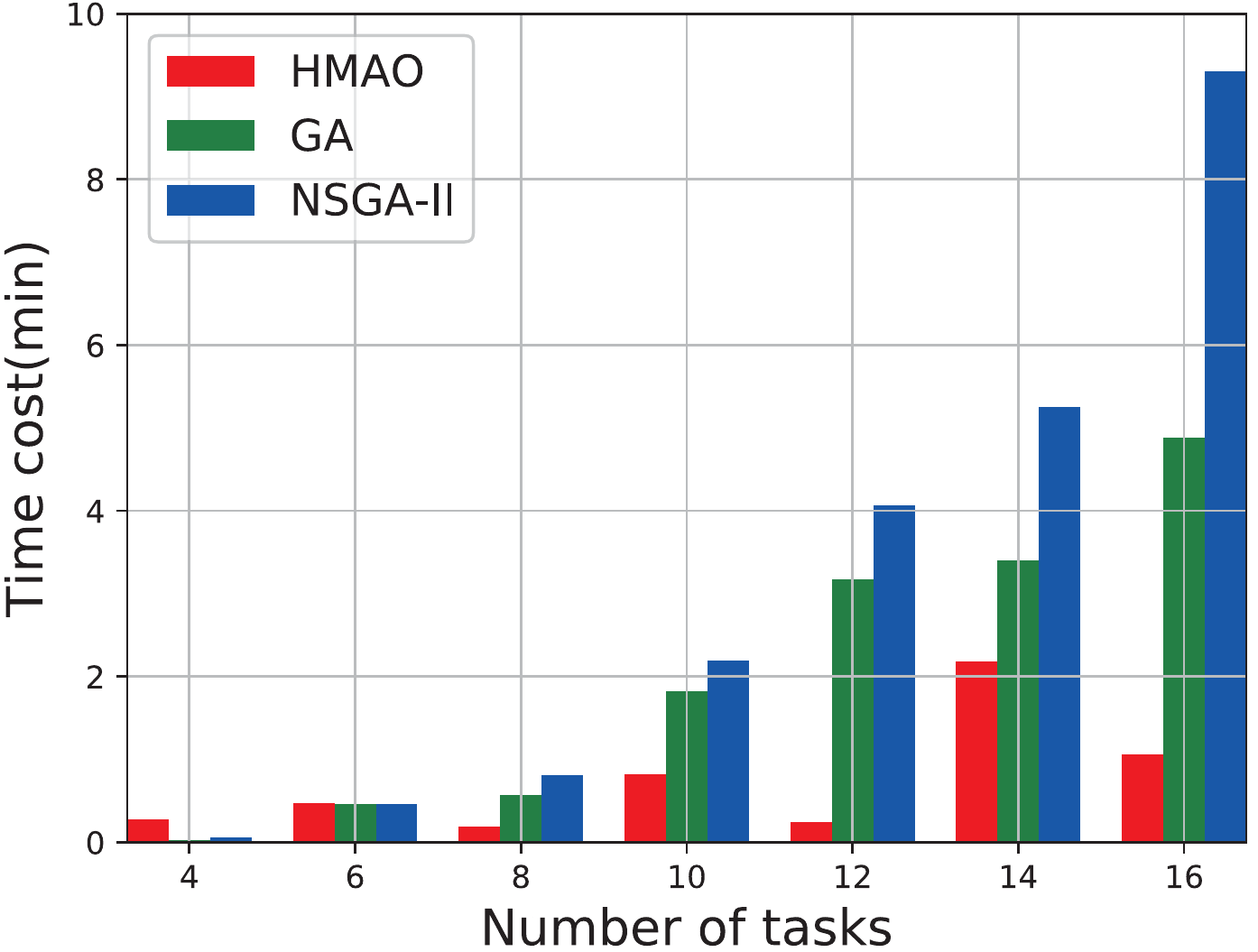}
  \label{Convergence time}}
  \caption{Performance comparisons for HMAO, GA and NSGA-II.}
  \label{Comparisons of different algorithms}
\end{figure}
In order to further analyze the performance of the proposed HMAO algorithm, we provide the maximum, minimum, mean, standard deviation and average convergence time of the simulation results for the proposed HMAO algorithm, GA and NSGA-II, and the detailed information can be observed from Table \ref{Results of simulation for HMAO, GA and NSGA-II}. In all test cases except $L=11$, the maximum, minimum, mean and standard deviation of the optimal solutions obtained by the proposed HMAO algorithm are better than or equal to GA and NSGA-II, respectively. Furthermore, the convergence time of the proposed HMAO algorithm is slower than GA, NSGA-II for $L=4,6$ but faster for the other cases, the difference is more evident with increasing the number of tasks. For example, in the case of $L=14$, we can observe that the performance of the proposed HMAO algorithm increases by $3.27\%$, and $3.44\%$, the convergence time reduces by $35.67\%$ and $58.38\%$ for GA and NSGA-II, respectively. For $L=16$, the mean results of the proposed HMAO algorithm improve by $4.26\%$ and $3.49\%$, the convergence time decreases by $78.23\%$ and $88.57\%$ for GA and NSGA-II, respectively. The results can demonstrate that the proposed HMAO algorithm is an effective optimization approach for resource allocation and has a better performance in terms of solution quality, robustness and convergence.\par

\begin{table*}[htbp]
\centering
\caption{Results of simulation for HMAO, GA and NSGA-II.}
\label{Results of simulation for HMAO, GA and NSGA-II}
\resizebox{\textwidth}{!}{%
\begin{tabular}{|c|c|c|c|c|c|c|c|c|c|c|c|c|c|c|c|}
\hline
\multirow{2}{*}{L} & \multicolumn{5}{c|}{HMAO}                     & \multicolumn{5}{c|}{GA}                       & \multicolumn{5}{c|}{NSGA-II}                     \\ \cline{2-16}
                   & Min    & Max    & Mean   & Std    & Time(min) & Min    & Max    & Mean   & Std    & Time(min) & Min    & Max    & Mean   & Std    & Time(min) \\ \hline
4                  & 0.7869 & 0.7869 & 0.7869 & 0.0000 & 0.2737    & 0.7869 & 0.7869 & 0.7869 & 0.0000 & 0.0231    & 0.7869 & 0.7869 & 0.7869 & 0.0000 & 0.0569    \\ \hline
5                  & 0.7718 & 0.7718 & 0.7718 & 0.0000 & 0.0520    & 0.7718 & 0.7718 & 0.7718 & 0.0000 & 0.0481    & 0.7603 & 0.7718 & 0.7706 & 0.0034 & 0.1408    \\ \hline
6                  & 0.8170 & 0.8170 & 0.8170 & 0.0000 & 0.4752    & 0.817  & 0.817  & 0.817  & 0.0000 & 0.4573    & 0.7628 & 0.8170 & 0.7953 & 0.0265 & 0.4666    \\ \hline
7                  & 0.7989 & 0.7989 & 0.7989 & 0.0000 & 0.0645    & 0.7989 & 0.7989 & 0.7989 & 0.0000 & 0.2841    & 0.7567 & 0.7989 & 0.7937 & 0.0126 & 1.0380    \\ \hline
8                  & 0.7869 & 0.7869 & 0.7869 & 0.0000 & 0.1903    & 0.7869 & 0.7869 & 0.7869 & 0.0000 & 0.5710    & 0.7869 & 0.7869 & 0.7869 & 0.0000 & 0.8126    \\ \hline
9                  & 0.8170 & 0.8170 & 0.8170 & 0.0000 & 0.5219    & 0.8170 & 0.8170 & 0.8170 & 0.0000 & 2.3043    & 0.7782 & 0.8170 & 0.7899 & 0.0177 & 2.1777    \\ \hline
10                 & 0.8041 & 0.8041 & 0.8041 & 0.0000 & 0.8174    & 0.7975 & 0.8041 & 0.8034 & 0.0020 & 1.8199    & 0.7718 & 0.8041 & 0.7976 & 0.0129 & 2.1903    \\ \hline
11                 & 0.7368 & 0.8102 & 0.7874 & 0.0283 & 0.8174    & 0.7887 & 0.7944 & 0.7938 & 0.0017 & 1.7007    & 0.7771 & 0.7944 & 0.7909 & 0.0052 & 2.3302    \\ \hline
12                 & 0.8170 & 0.8170 & 0.8170 & 0.0000 & 0.2446    & 0.7818 & 0.8170 & 0.8045 & 0.0135 & 3.1743    & 0.7628 & 0.8170 & 0.7929 & 0.0161 & 4.0658    \\ \hline
13                 & 0.8070 & 0.8070 & 0.8070 & 0.0000 & 1.0464    & 0.7715 & 0.8069 & 0.7973 & 0.0104 & 3.8402    & 0.7762 & 0.8070 & 0.7977 & 0.0131 & 4.2793    \\ \hline
14                 & 0.8116 & 0.8119 & 0.8118 & 0.0001 & 2.1852    & 0.7677 & 0.7989 & 0.7861 & 0.0127 & 3.3968    & 0.7675 & 0.7989 & 0.7848 & 0.0119 & 5.2505    \\ \hline
15                 & 0.8170 & 0.8170 & 0.8170 & 0.0000 & 0.9704    & 0.7590 & 0.7923 & 0.7811 & 0.0099 & 4.3592    & 0.7680 & 0.8075 & 0.7875 & 0.0115 & 7.7129    \\ \hline
16                 & 0.8088 & 0.8088 & 0.8088 & 0.0000 & 1.0624    & 0.7576 & 0.7869 & 0.7757 & 0.0092 & 4.8812    & 0.7714 & 0.7868 & 0.7815 & 0.0049 & 9.3008    \\ \hline
\end{tabular}%
}
\end{table*}

Furthermore, we compare the results of the proposed HMAO algorithm with GA and NSGA-II in Fig. \ref{Comparisons of different algorithms}, including resource and bandwidth utilization, optimal solution and average convergence time. Fig. \ref{Resource utilization} shows the resource utilization for different tasks, the results obtained by the proposed HMAO algorithm are equal to GA for $L=4,6,8,10$, NSGA-II for $L=4,8$, and better than GA and NSGA-II for the other cases. The solution with a lower resource utilization implies that there are more service nodes to be used to deploy the requested tasks. The results of bandwidth utilization are described in Fig. \ref{Bandwidth utilization}. We can observe that the performance of the proposed HMAO algorithm is better than or equal to GA and NSGA-II for $L=4,6,8,10$. For $L=12,14,16$, it is due to the fact that the number of service nodes for the proposed HMAO algorithm is less than that of the other two baseline algorithms. Fig. \ref{Optimal solution} illustrates the results of optimal solutions, the performance for the proposed HMAO algorithm, GA and NSGA-II is approximately equivalent with a lower number of tasks, and the proposed HMAO algorithm performs better than GA and NSGA-II as the number of tasks increases. The average convergence time of three algorithms can be observed from Fig. \ref{Convergence time}, the time taken to converge by the proposed HMAO algorithm is worse than that of GA and NSGA-II as the number of tasks is small, but the proposed HMAO algorithm outperforms GA and NSGA-II as the number of tasks increases.\par

\subsection{Evaluate HMAO in On-line Resource Allocation}\label{Performance Analysis of the HMAO in On-line Resource Allocation}

\begin{figure}[tbp]
  \centering
  \subfigure[Resource utilization with $L_0$]{\includegraphics[width=0.23\textwidth]{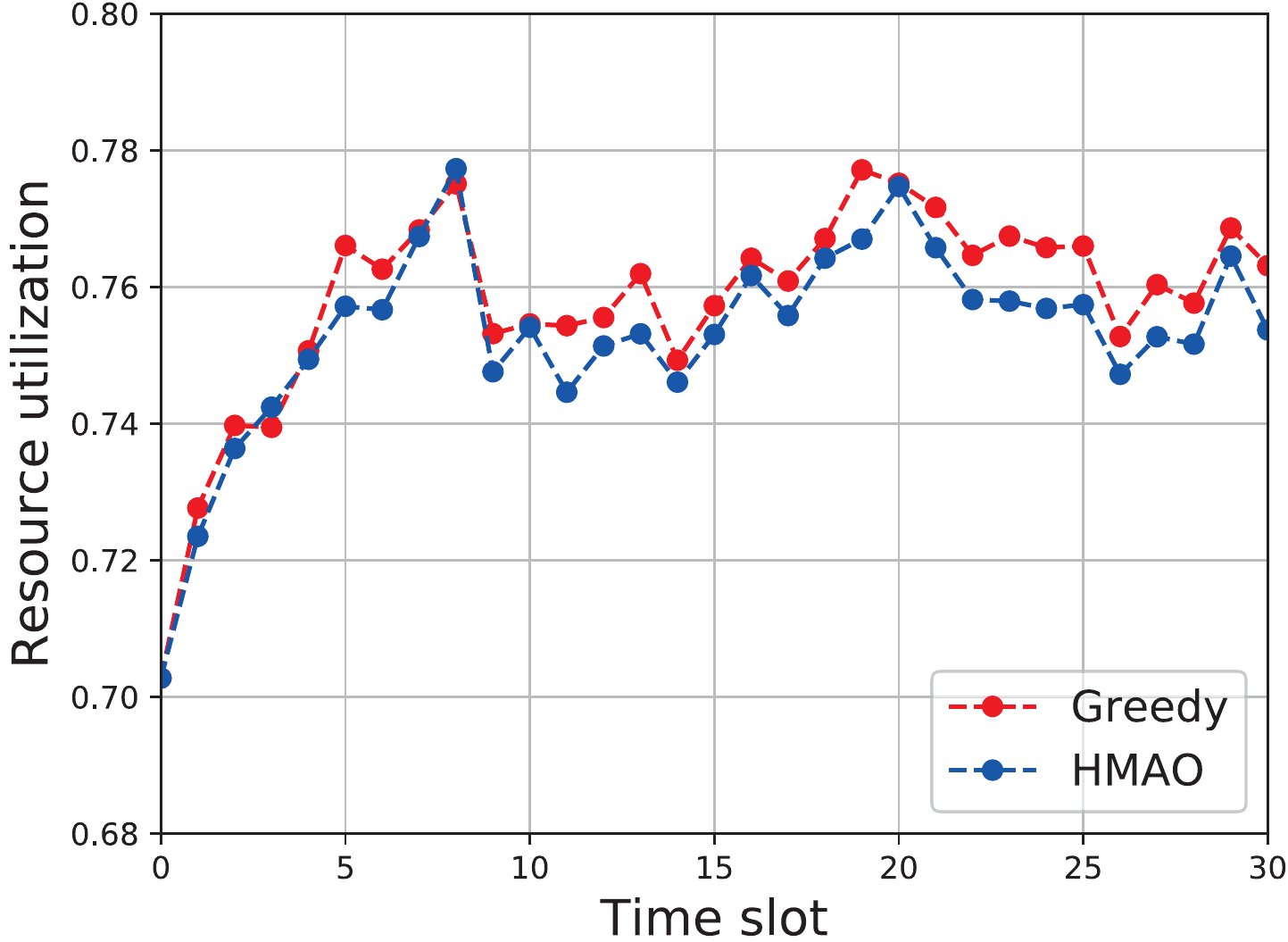}
  \label{Resource utilization in dynamical environment}}
  \subfigure[Bandwidth utilization with $L_0$]{\includegraphics[width=0.23\textwidth]{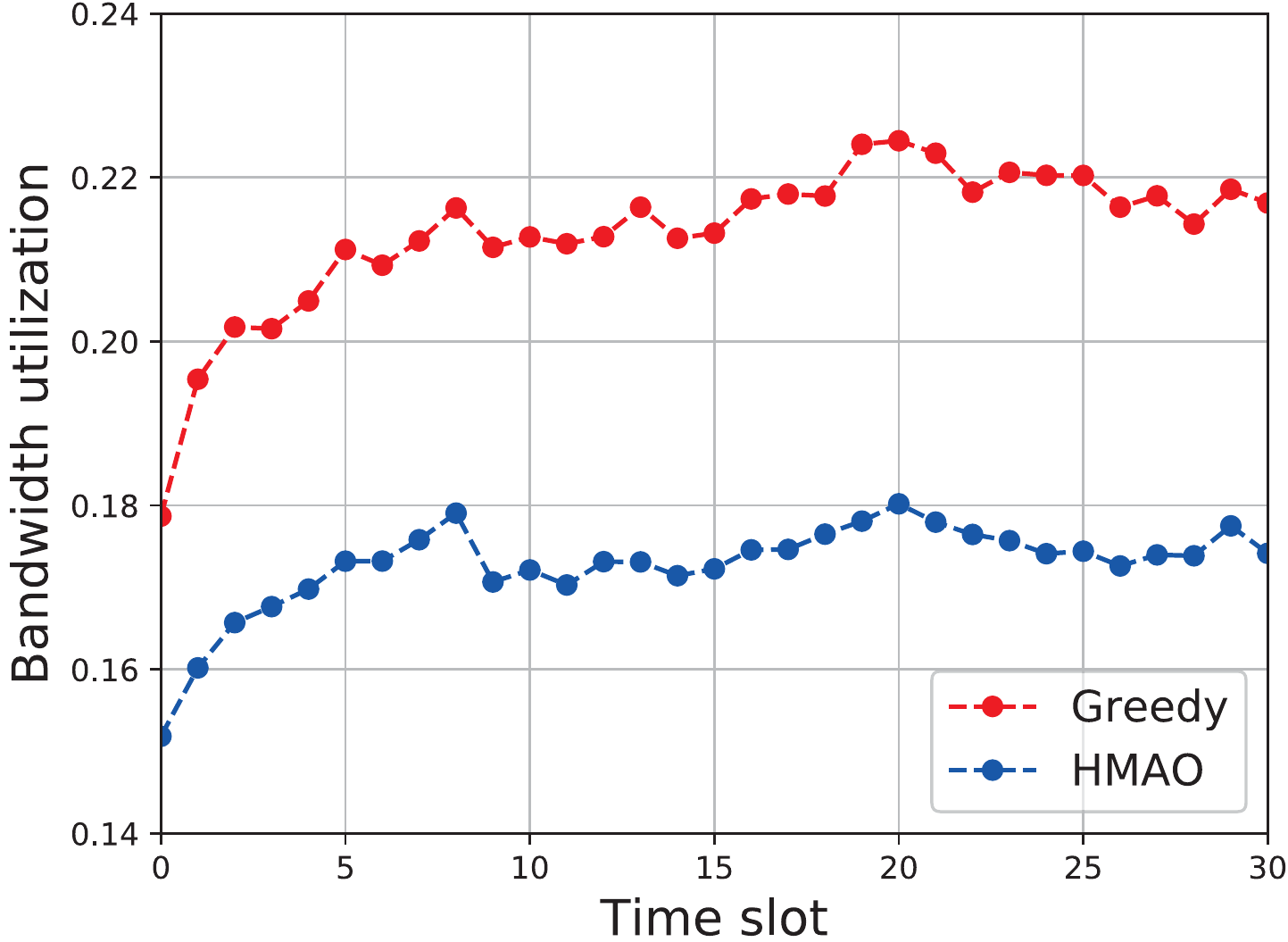}
  \label{Bandwidth utilization in dynamical environment}}
  \subfigure[Optimal solution with $L_0$]{\includegraphics[width=0.23\textwidth]{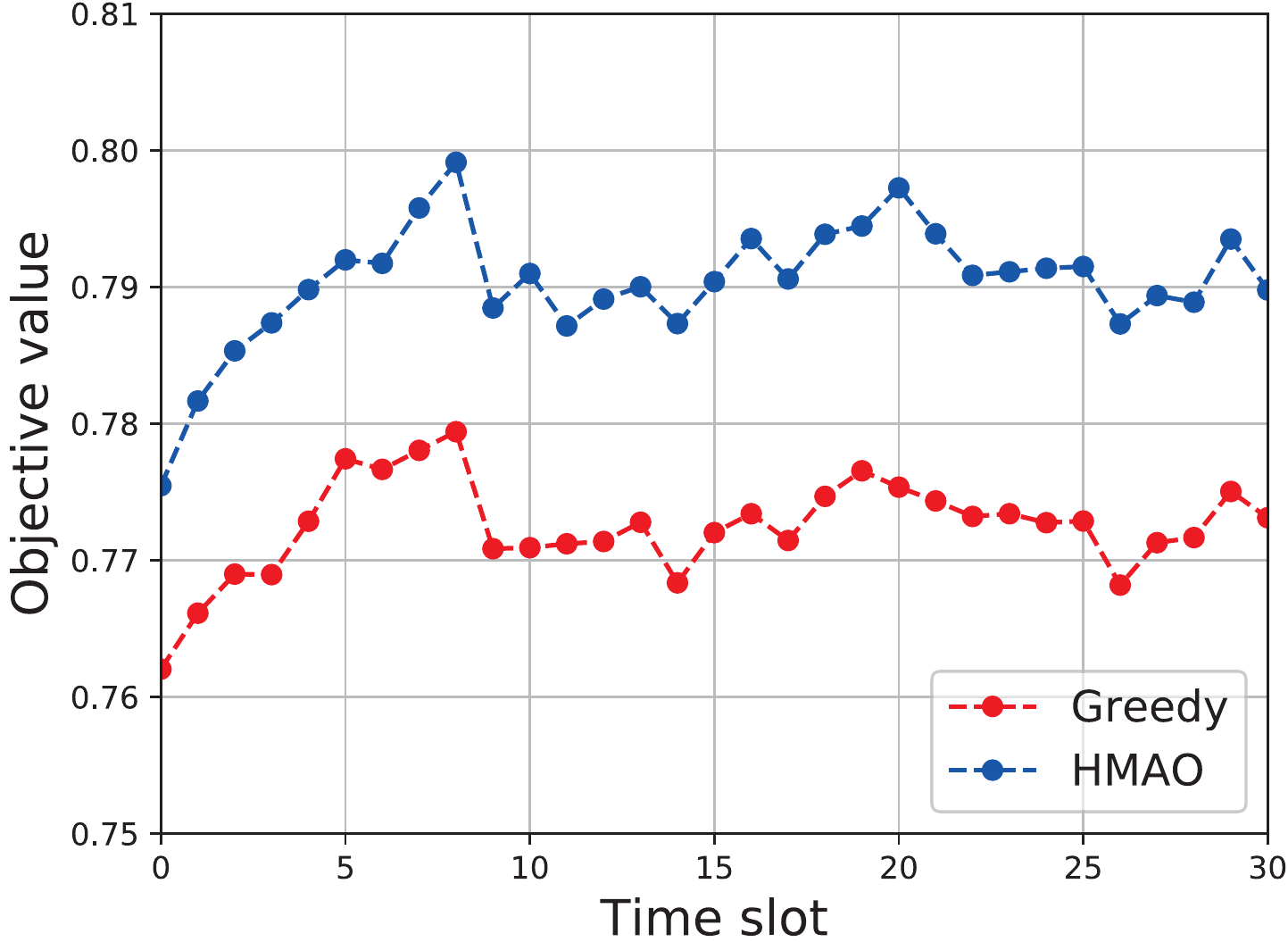}
  \label{Optimal solution in dynamical environment}}
  \subfigure[Optimal solutions with $L_{0\sim 10}$]{\includegraphics[width=0.23\textwidth]{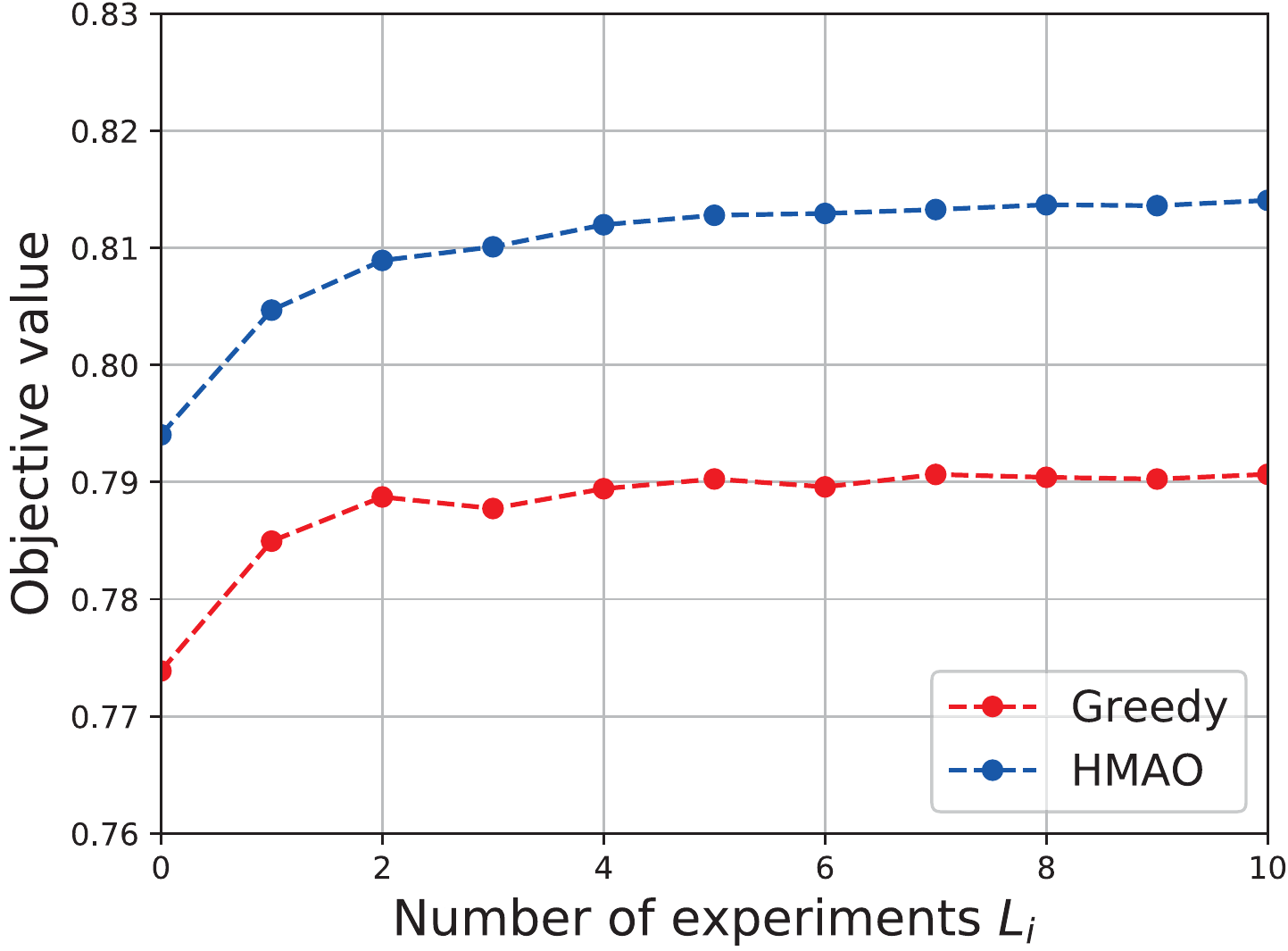}
  \label{Optimal solution with all cases in dynamical environment}}
  \caption{Performance comparisons for HMAO and Greedy.}
  \label{Performance of the HMAO in dynamical environment}
\end{figure}

To further investigate the performance of the proposed HMAO algorithm in on-line resource allocation, we implement the resource allocation model in a dynamic environment and design several experiments of dynamically allocating the available resources to the requested tasks with the parameters given in Table \ref{Parameter setting for HMAO, GA and NSGA-II}. We assume that there are new requested tasks are appearing and old requested tasks are ending in each time slot, whose numbers are randomly produced from $L_i = \left \{ 4*i+1,\dots, 4*i+9 \right \}, i=0,1,\dots,10$. Each case also runs 10 times with time slots from 0 to 30 and the average results of optimal solutions are obtained by the proposed HMAO algorithm. Moreover, we compare the proposed HMAO algorithm with a first-fit greedy approach \cite{Espling2016modeling} in terms of resource utilization, bandwidth utilization and optimal solution. The simulation results for the case $L_0$ can be observed from Fig. \ref{Performance of the HMAO in dynamical environment}, where we provide the performance of optimal solutions for all cases $L_0 \sim L_{10}$ by the proposed HMAO algorithm and the greedy approach in Fig. \ref{Optimal solution with all cases in dynamical environment}.\par

Fig. \ref{Resource utilization in dynamical environment} shows the computing resource utilization for $L_0$. It can be observed that the number of computing resources are comparatively close between the proposed HMAO algorithm and the greedy approach in each time slot. The results of bandwidth utilization for $L_0$ are illustrated in Fig. \ref{Bandwidth utilization in dynamical environment}, where we can observe that the bandwidth utilization values obtained by the proposed HMAO algorithm are better than that of the greedy approach for all time slots, and there is average $19.00\%$ performance improvement. This is because that the adjacent sub-tasks of a task can be migrated and swapped to the same service node by the proposed HMAO algorithm as far as possible in order to reduce the bandwidth resources used by the requested tasks. Fig. \ref{Optimal solution in dynamical environment} describes the objective results of optimal solution for $L_0$, where we can observe that the proposed HMAO algorithm performs better than the greedy approach and the performance of the proposed HMAO algorithm shows an average increase of $2.31\%$ when compared with the greedy approach. To better analyze the performance of the proposed HMAO algorithm for processing the different number of the requested tasks in a time slot, we provide the average results of optimal solutions for $L_0 \sim L_{10}$ in Fig. \ref{Optimal solution with all cases in dynamical environment}. It is obvious that the proposed HMAO algorithm outperforms the greedy approach for $L_0 \sim L_{10}$, and the average performance improvement of the proposed HMAO algorithm is nearly $2.81\%$. From Fig. \ref{Performance of the HMAO in dynamical environment}, we can observe that the proposed HMAO algorithm is better than the first-fit greedy approach in dynamical resource allocation.

\section{Conclusion}\label{Conclusion}

This paper studies the problem of resource allocation in cloud computing systems. Our aim is to maximize the resource utilization based on CPU, memory and GPU, and minimize the bandwidth cost. To address the problem, we propose the HMAO algorithm which combines the improved GA and the MAO algorithm, where the improved GA is to find an optimal resource utilization solution and the MAO algorithm is to minimize the bandwidth cost. For the MAO algorithm, we use a priority-based selection mechanism to obtain the candidate source sub-tasks, and design the selection and exchange operators by a probabilistic method to migrate and swap the sub-tasks on several service agents. The proposed HMAO algorithm can obtain the objective optimal result by cooperative co-evolutionary method.\par

Finally, we verify and evaluate the performance of the proposed HMAO algorithm via simulation experiments. When compared with GA and NSGA-II, we can observe that the proposed HMAO algorithm outperforms them in terms of solution quality, convergence time and robustness as the number of tasks increases. For $L=14$, the proposed HMAO algorithm improves performance by $3.38\%$ for GA and $3.44\%$ for NSGA-II, and reduces the average convergence time by $19.34\%$ for GA and $58.38\%$ for NSGA-II. Furthermore, we compare the performance of the proposed HMAO algorithm with the first-fit greedy approach in on-line resource allocation. We can observe that the performance of the proposed HMAO algorithm increases approximately by $2.81\%$. \par

\ifCLASSOPTIONcaptionsoff
  \newpage
\fi



%
%
%
\bibliographystyle{IEEEtran}
\bibliography{bibfile}
%

\begin{IEEEbiography}[{\includegraphics[width=1in,height=1.25in,clip,keepaspectratio]{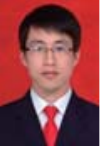}}]{Xiangqiang Gao}
Xiangqiang Gao received the B.Sc. degree in school of electronic engineering from Xidian University and the M.Sc. degree from Xi\textquoteright an Microelectrinics Technology Institute, Xi\textquoteright an, China, in 2012 and 2015, respectively. He is currently pursuing the Ph.D. degree with the School of Electronic and Information Engineering, Beihang University, Beijing, China. His research interests include rateless codes, software defined network and network function virtualization.\par
\end{IEEEbiography}

\begin{IEEEbiography}[{\includegraphics[width=1in,height=1.25in,clip,keepaspectratio]{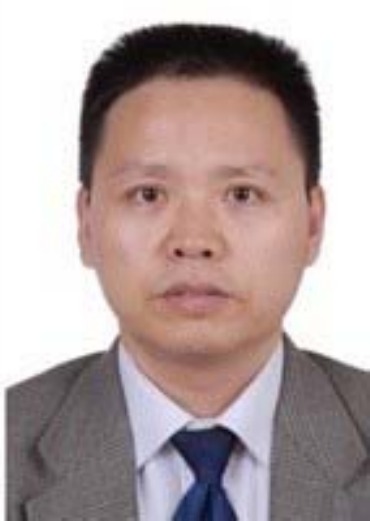}}]{Rongke Liu}
Rongke Liu received the B.Sc. degree in electronic engineering and Ph.D. degree in information and communication engineering from Beihang University, Beijing, China, in 1996 and 2002, respectively. From 2006 to 2007, he was a visiting professor at Florida Institute of Technology, Florida. In August, 2015, he visited the university of Tokyo as a senior visiting scholar. He is a Full Professor with the School of Electronic and Information Engineering in Beihang University, specializing in the fields of information and communication engineering. He has authored or co-authored more than 100 papers in journals and conferences, and edited four books. His current research interests include multimedia computing and space information network. He is a Member of the IEEE and ACM. Dr. Liu was one of the winners of education ministry\textquoteright s New Century Excellent Talents supporting plan in 2012.\par
\end{IEEEbiography}
\vfill
\newpage
\begin{IEEEbiography}[{\includegraphics[width=1in,height=1.25in,clip,keepaspectratio]{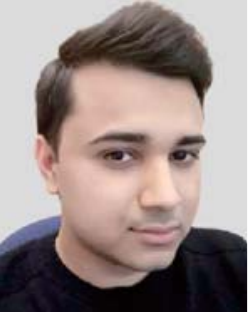}}]{Aryan Kaushik}
Aryan Kaushik received the M.Sc. degree in telecommunications from the Hong Kong University of Science and Technology, Hong Kong, in 2015. He is currently pursuing the Ph.D. degree in communications engineering with the Institute for Digital Communications, University of Edinburgh, U.K., where  he is also pursuing the postgraduate certification in academic practice with the Institute for Academic Development. He has been a Visiting Researcher with Imperial College London, U.K., in 2019; the University of Luxembourg, Luxembourg, in 2018; and Beihang University, China, in 2017 and 2018. His research interests are in the area of signal processing for communications, green wireless communications for 5G and beyond, and millimeter wave multi antenna systems.\par
\end{IEEEbiography}

\vfill


\end{document}